\documentclass[journal,onecolumn,web]{ieeecolor}
\usepackage{generic}
\usepackage{cite}
\usepackage{amsmath,amssymb,amsfonts,dsfont,mathrsfs}
\usepackage[ruled,boxed]{algorithm2e}
\usepackage{graphicx}
\usepackage{subfigure}
\usepackage{multirow}
\usepackage{cases}
\usepackage{float}
\usepackage{setspace}
\usepackage{slashbox}
\usepackage{boxedminipage}
\usepackage{pdfsync}
\usepackage{empheq}
\usepackage{xcolor}
\definecolor{mycyan}{HTML}{0070C0}
\usepackage[colorlinks,linkcolor=mycyan,anchorcolor=black,citecolor=mycyan,urlcolor=mycyan]{hyperref}

\newcommand{\bassume}{ \begin{assume} \begin{rm} }
\newcommand{\eassume}{ \end{rm} \hfill $\triangleleft$ \end{assume} }
\newcommand{\bcondition}{ \begin{condition} \begin{rm} }
\newcommand{\econdition}{ \end{rm}  \end{condition} }
\newcommand{\bremark}{ \begin{remark} \begin{rm} }
\newcommand{\eremark}{ \end{rm} \hfill $\triangleleft$ \end{remark} }
\newcommand{\btheorem}{ \begin{theorem} \begin{rm} }
\newcommand{\etheorem}{ \end{rm} \hfill $\triangleleft$ \end{theorem} }
\newcommand{\blemma}{ \begin{lemma} \begin{rm} }
\newcommand{\elemma}{ \end{rm} \hfill $\triangleleft$ \end{lemma} }
\newcommand{\bcorollary}{ \begin{corollary} \begin{rm} }
\newcommand{\ecorollary}{ \end{rm}  \end{corollary} }
\newcommand{\bdefinition}{ \begin{definition}\begin{rm} }
\newcommand{\edefinition}{ \end{rm} \hfill $\triangleleft$ \end{definition} }
\newcommand{\bproposition}{ \begin{proposition} \begin{rm} }
\newcommand{\eproposition}{ \end{rm}  \end{proposition} }
\newcommand{\bexample}{ \begin{example} \begin{rm} }
\newcommand{\eexample}{ \end{rm} \hfill $\triangleleft$ \end{example} }
\newcommand{\bproblem}{ \bf Problem \begin{rm} }
\newcommand{\eproblem}{ \end{rm} \hfill $\triangleleft$ \end{problem} }
\newcommand{\bproof}{ \textit{Proof:} \begin{rm} }
\newcommand{\eproof}{ \end{rm} \hfill $\square$}
\newtheorem{theorem}{\it Theorem}
\newtheorem{lemma}{\it Lemma}
\newtheorem{definition}{\it Definition}
\newtheorem{remark}{\it Remark}
\newtheorem{corollary}{\it Corollary}
\newtheorem{proposition}{\it Proposition}
\newtheorem{example}{\it Example}
\newtheorem{assume}{\it Assumption}
\newtheorem{condition}{\it Condition}

\def\BibTeX{{\rm B\kern-.05em{\sc i\kern-.025em b}\kern-.08em
    T\kern-.1667em\lower.7ex\hbox{E}\kern-.125emX}}
\begin{document}
\title{Robust Mean Field Social Control: A Unified Reinforcement Learning Framework}
\author{
 Zhenhui Xu, Jiayu Chen, Bing-Chang Wang, Yuhu Wu, and Tielong Shen
% \thanks{This paragraph of the first footnote will contain the date on
% which you submitted your paper for review. It will also contain support
% information, including sponsor and financial support acknowledgment. For
% example, ``This work was supported in part by the U.S. Department of
% Commerce under Grant BS123456.'' }
% \thanks{The next few paragraphs should contain
% the authors' current affiliations, including current address and e-mail. For
% example, F. A. Author is with the National Institute of Standards and
% Technology, Boulder, CO 80305 USA (e-mail: author@boulder.nist.gov). }
\thanks{Zhenhui Xu and Jiayu Chen are with the Department of Systems and Control Engineering at the Institute of Science Tokyo, Tokyo, 152-8552, Japan. (e-mail: xuzhenhui@eagle.sophia.ac.jp;jiayuc@mail.ustc.edu.cn).}
\thanks{Bing-Chang Wang is with the School of Control Science and Engineering, Shandong University, Jinan, 250061, China. (e-mail: bcwang@sdu.edu.cn)}
\thanks{Yuhu Wu and Tielong Shen are with the School of Control Science and Engineering, Dalian University of Technology, Dalian, 116024, China (e-mail: wuyuhu@dlut.edu.cn;tetu-sin@sophia.ac.jp).}}

\maketitle

\begin{abstract}
This paper studies linear quadratic Gaussian robust mean field social control problems in the presence of multiplicative noise. We aim to compute asymptotic decentralized strategies without requiring full prior knowledge of agents’ dynamics. The primary challenges lie in solving an indefinite stochastic algebraic Riccati equation for feedback gains, and an indefinite algebraic Riccati equation for feedforward gains. To overcome these challenges, we first propose a unified dual-loop iterative framework that handles both indefinite Riccati-type equations, and provide rigorous convergence proofs for both the outer-loop and inner-loop iterations. Secondly, considering the potential biases arising in the iterative processes due to estimation and modeling errors, we verify the robustness of the proposed algorithm using the small-disturbance input-to-state stability technique. Convergence to a neighborhood of the optimal solution is thus ensured, even in the existence of disturbances. Finally, to relax the limitation of requiring precise knowledge of agents' dynamics, we employ the integral reinforcement learning technique to develop a data-driven method within the dual-loop iterative framework. A numerical example is provided to demonstrate the effectiveness of the proposed algorithm.
\end{abstract}
\begin{IEEEkeywords}
Robust mean field social control, indefinite SARE, data-driven control, reinforcement learning, robustness
\end{IEEEkeywords}

\section{Introduction}
\label{sec:introduction}
Mean field game (MFG) theory, introduced independently by Lasry and Lions \cite{lasry2007mean} and Huang, Caines, and Malham{\'e} \cite{huang2007large}, provides a powerful framework for analyzing stochastic differential games involving a large number of interacting agents. In an MFG setting, each agent seeks to optimize its own cost functional while being influenced by a mean field term that represents the aggregate effect of all agents. This approach effectively decouples individual interactions, leading to a mean field equilibrium characterized by tractable forward-backward equations: the Kolmogorov-Fokker-Planck equation describes the evolution of the mean field distribution, and the Hamilton-Jacobi-Bellman equation governs the value function for each agent \cite{cousin2011mean,bensoussan2013mean,carmona2018probabilistic}. Mean field social control (MFSC) extends the MFG framework from individual optimization to collective optimization, where the objective is to design decentralized strategies that minimize a common social cost, typically defined as the sum or average of all agents' costs. Two main approaches are commonly used: the person-by-person optimization approach \cite{bauso2016opinion,huang2016linear}, which reformulates the cooperative problem as a non-cooperative game and applies MFG tools to find the social optimum; and the direct approach \cite{huang2021linear2,wang2020indefinite}, which starts by directly solving a cooperative game involving all agents and then analyzes the resulting coupled equations in the limit as the number of agents tends to infinity. \textcolor{black}{A fundamental question in both MFG and MFSC is whether the mean field solution provides an accurate approximation of the corresponding finite-agent problem. In contrast, research on mean field type control problems has mainly focused on the existence and uniqueness of mean field solutions, which have been widely extended to the development of multi-agent learning methods \cite{lauriere2014dynamic,gu2023dynamic,mondal2022approximation,li2024policy,zaman2024robust}. In this setting, the limit optimal control problem is formulated for a representative agent, where the dynamics of the agent's state and the cost functional depend on the probability distribution of the state. Notably, all mean field formulations reduce high-dimensional multi-agent problems to tractable low-dimensional PDE systems, and have been applied in various fields \cite{aurell2022optimal,wang2024mean,kizilkale2019integral}.} 

Despite these advances, environmental uncertainties in practical systems often render the idealized assumptions of classical analyses invalid. To address this challenge, robust mean field formulations have been developed to provide strategies that are resilient to model mismatches and external disturbances. Significant progress has been made, especially in the linear quadratic Gaussian (LQG) setting, due to its analytical tractability and broad applicability \cite{huang2013mean,tembine2013robust,moon2016linear,huang2017robust,wang2020social,liang2022robust}. In this context, uncertainty is often induced by common fluctuating factors, such as taxation, subsidy, or interest rate, and is modeled as an adversarial input evaluated under a worst-case hypothesis. Within this framework, robust MFSC problems have been studied under various noise structures. For example, \cite{wang2020social} studied the additive noise case by formulating the problem as two team-zero-sum games involving both control and disturbance inputs for all agents. The objective is to attenuate the effect of disturbances on the social cost by rejecting $L_2$-gain from the disturbance to the performance-related signal of each agent. This leads to asymptotic decentralized strategies derived from solving Riccati-type equations with indefinite terms, where the indefiniteness stems from the disturbance channel. More recently, \cite{liang2022robust} extended the analysis to the multiplicative noise case, and characterized the robust mean field social optimum by low-dimensional indefinite stochastic Riccati-type equations coupled with ordinary differential equations (ODEs), which represent the LQG counterpart of forward-backward PDF systems encountered in more general mean field models. 

\textcolor{black}{Consequently, solving indefinite Riccati  equations is essential for establishing mean field social optima in LQG–MFSC problems. These equations also arise in the context of $H_{\infty}$ control problems for single-agent systems \cite{bacsar2008h,zhou1996robust}. In particular, the algebraic Riccati equation (ARE) with an indefinite quadratic term is associated with deterministic systems and stochastic systems with additive noise in the continuous-time infinite-horizon setting, while the indefinite stochastic
algebraic Riccati equation (SARE) arises in stochastic systems with multiplicative noise. These problems are significantly more challenging than their definite counterparts  since the presence of indefinite quadratic terms prevents the direct application of standard convex optimization techniques.}

\textcolor{black}{The complexity of indefinite Riccati type equations has led to the development of numerical solution methods, in which reinforcement learning (RL) has recently played an increasingly important role. As a prominent branch of machine learning, RL enables agents to learn optimal strategies through iterative interactions with their environment, guided by feedback in the form of rewards or penalties \cite{sutton2018reinforcement,vamvoudakis2010online,mondal2022approximation,zaman2024robust}. 
In the field of optimal control, one of the fundamental RL approaches is policy iteration (PI) (see \cite{kleinman1968iterative,li2022stochastic}), where control policies are successively improved through policy evaluation and policy improvement steps. Building on this foundation, integral reinforcement learning (IRL) was developed to eliminate the need for explicit system models in the iterative procedure \cite{vamvoudakis2010online,modares2014optimal,song2016off}. 
Several iterative approaches, notably those based on IRL, have been proposed for indefinite AREs. In particular, \cite{lanzon2008computing} transforms the problem into solving a sequence of $H_2$-type AREs, the quadratic terms of which are negative semidefinite. It has also been proven that this sequence converges to the stabilizing solution of the indefinite ARE. Building on  this framework, \cite{vrabie2011adaptive} introduced a {\color{black}dual-loop IRL scheme} in which control and disturbance inputs are updated in separate loops. However, a complete convergence analysis was not provided. A single-loop alternative was proposed in \cite{wu2013simultaneous}, where both inputs are updated simultaneously, and local convergence was established via a Newton sequence \cite{rall1974note}. A rigorous convergence proof was later given in \cite{liu2019new} under additional conditions by introducing an initialization criterion based on Kantorovich’s theorem. For the indefinite SARE, \cite{dragan2011computation} approximates the stabilizing solution via a sequence of $H_2$-type SAREs with negative semidefinite quadratic terms, though the state weighting matrix may still be indefinite. In the special case of an $H_2$-type SARE with a positive state weighting matrix, \cite{li2022stochastic} developed a PI method enhanced by IRL to reduce model dependence. \cite{xu2024mean} extended this method to a fully model-free setting.} 

\subsection{Motivation and contributions}
Although recent advances in robust MFSC problems have provided important theoretical insights, most studies have focused on characterizing robust mean field social optima rather than developing efficient numerical methods. Practical applications, however, demand tractable algorithms, especially when agents’ dynamics are only partially known or completely unknown. \textcolor{black}{Motivated by this gap, we investigate the robust LQG–MFSC problem with multiplicative noise in the dynamics and population coupling in the cost functional. This model is considered out of two reasons. First, unlike additive noise, multiplicative noise captures situations where the noise intensity depends on both the state and the inputs. This leads to a richer and more realistic modeling framework where variability scales with activity levels ({\sl e.g.}, volatility in finance). Second, in many practical scenarios, such as epidemic control \cite{aurell2022optimal}, wireless power control \cite{huang2003individual}, and smart-grid load coordination \cite{kizilkale2019integral}, agents possess inherently individual dynamics while interacting through the cost functional.}

\textcolor{black}{Recently, \cite{zaman2024robust} studied a related robust mean field problem in the discrete time and additive noise setting. They developed a receding-horizon gradient descent ascent algorithm that takes the form of a dual-loop iterative framework. However, their method cannot be applied directly to our setting since the outer-loop update proceeds backward in time from a finite horizon. Our problem, by contrary, is formulated on an infinite horizon, which prevents such a backward recursion. In our case, the direct approach \cite{liang2022robust} reduces the task of finding the mean field social optimum to solving three key equations: an indefinite SARE, an indefinite ARE, and an ordinary differential equation (ODE). The indefinite SARE, in particular, presents challenges that are not addressed by existing methods. To resolve this, we propose a unified dual-loop iterative framework for both indefinite SARE and ARE.}

\textcolor{black}{Our approach tackles three key technical hurdles:} 
1). It is intractable to apply the PI algorithm directly to the $H_2$-type SARE framework proposed in \cite{dragan2011computation}. To address this issue, we construct a new sequence of $H_2$-type SAREs in the outer loop and prove convergence. Additionally, the convergence results of traditional PI methods for (S)AREs rely on positive (semi)definite state weighting matrices, however, our generated equations may involve indefinite state weighting matrices. We therefore develop a novel convergence analysis tailored to this case, which broadens the applicability of PI algorithms.
2). Beyond disturbances in agents’ dynamics, iterative processes also suffer from biases caused by estimation and modeling errors. To address this issue, we model the inexact iterations as discrete-time systems and treat the errors as external disturbances. Using the small-disturbance input-to-state stability (ISS) framework (\cite{pang2021robust,cui2024robust}), we prove that both the inner-loop and outer-loop iterative processes remain stable, even in the presence of small disturbances. 3). The transition from a model-based iterative scheme to a data-driven approach is difficult, particularly in  for the indefinite ARE. First, this equation includes a state weighting matrix that depend on the system parameters. We introduce an equation transformation to eliminate these direct parameter dependencies. Second, the presence of multiplicative noise hinders the direct application of IRL via agent trajectories. We transform the original dynamics into a deterministic one, which facilitates leveraging variations in the agents’ trajectories.  
 
Our main contributions are summarized as follows: (\romannumeral 1) A novel unified dual-loop iterative framework is proposed for solving both indefinite SARE and ARE, with guaranteed convergence. (\romannumeral 2) Robustness analyses of both the outer-loop and inner-loop iterations are conducted using the small-disturbance ISS technique, ensuring the convergence in the presence of disturbances. (\romannumeral 3) A novel data-driven algorithm is developed to approximate the mean field social optimum strategies without requiring full prior knowledge of agents' dynamics.

\subsection{Organization and Notations}
This paper is organized as follows. Section \ref{sec2} describes the problem formulation and presents basic results for robust MFSC. In section \ref{sec:main result1}, a dual-loop iterative framework for solving indefinite Riccati-type equations is introduced, along with rigorous proofs of convergence. Section \ref{sec:main result2} provides theoretical analyses of the robustness properties for both iterative processes. In Section \ref{sec:main result3}, the proposed framework is applied to solve the robust MFSC problem, and a data-driven method is developed. Section \ref{sec:simulation} presents a simulation example to demonstrate the effectiveness of the proposed approach. Finally, Section \ref{sec:conclusion} concludes the paper.
 
% $\|\cdot\|_2$ and $\|\cdot\|_F$ represent the spectral norm and the Frobenius norm of a matrix, respectively.
A list of notations is presented as follows. For a family of $\mathbb{R}^n$-valued random variables $\{x(\tau),\tau\geq0\}$, $\sigma(x(\tau),\tau\leq t)$ is the $\sigma$-algebra generated by these random variables. \textcolor{black}{For a matrix $A\in\mathbb{R}^{n\times m}$ and $Q\in\mathbb{R}^{n\times n}$, $|A|_Q^2=A^{\top}QA$. In particular, when $Q=I_n$, the subscript is omitted.} $\mathrm{vec}(A)$ denotes the column-wise vectorization of $A$ into a vector $\mathbb{R}^{mn}$. For a symmetric matrix $P\in\mathbb{S}^n$, $\mathrm{vecm}(P)$ denotes the vectorization with only its upper triangular part, {\sl i.e.,} $[p_{11},p_{12},\cdots,p_{1n},p_{22},p_{23},\cdots,p_{nn}]^{\mathrm{T}}\in\mathbb{R}^{\frac{n(n+1)}{2}}$. For a family of $\mathbb{R}^n$-valued random variables $\{x(\tau),\tau\geq0\}$ and $\mathbb{R}^m$-valued random variables $\{y(\tau),\tau\geq0\}$, and a given time interval $T>0$, we define: \textcolor{black}{$\delta_{x}^t = \mathrm{vecm}(2x(t+T)x(t+T)^{\top}-\mathrm{diag}(x(t+T))^{2})-\mathrm{vecm}(2x(t)x(t)^{\top}-\mathrm{diag}(x(t))^{2})$, $\delta_{xy}^t=x(t+T)\otimes y(t+T)-x(t)\otimes y(t)$, $I_{x}^{t}=\int_{t}^{t+T}\mathrm{vecm}(2x(\tau)x(\tau)^{\top}-\mathrm{diag}(x(\tau))^2)\mathrm{d}\tau$, $I_{xy}^t=\int_{t}^{t+T}x(\tau)\otimes y(\tau)\mathrm{d}\tau$.} \textcolor{black}{Let $A^{\dag} = (A^{\top}A)^{-1}A^{\top}$
denote the Moore-Penrose inverse of a full column-rank
matrix $A$. The symbols $\|\cdot\|_F$ and $\|\cdot\|_2$ denote the Frobenius and spectral norms, respectively.}

\section{Problem formulation}\label{sec2}
 Consider a population of $N$ agents, denoted by $\mathcal{A}=\{\mathcal{A}_1,\cdots,\mathcal{A}_N\}$, where $\mathcal{A}_i$ represents the $i$-th agent. The state process $x_i(t)\in\mathbb{R}^n$ of $\mathcal{A}_i$ evolves according to the following stochastic differential equation (SDE)
\begin{equation}\label{sys1}
\left\{\begin{aligned}
&\mathrm{d}x_i(t)=\left[Ax_i(t)+Bu_i(t)+Gv_i(t)\right]\mathrm{d}t\\
&~~~~~~~~~~~+\left[Cx_i(t) + Du_i(t)\right]\mathrm{d}w_i(t),\\
&x_i(0) =x_{i0}, 
\end{aligned}\right.
\end{equation}
where $u_i\in\mathbb{R}^{m_1}$, and $v_i\in\mathbb{R}^{m_2}$ represent the control and disturbance inputs for $\mathcal{A}_i$, respectively. $\{w_i(t),1\leq i\leq N\}$ are a sequence of independent one-dimensional Brownian motions defined on a complete filtered probability space $(\Omega,\mathcal{F},\{\mathcal{F}_t\}_{t\geq0},\mathbb{P})$. The initial states $\{x_{i0},1\leq i\leq N\}$ are mutually independent and have the same expectation ({\sl i.e.}, $\mathbb{E}[x_{i0}]=\bar{x}_0$) and a finite second moment. These initial states are also independent of $\{w_i(t),1\leq i\leq N\}$. The matrices $A,B,G,C,D$ are constant with compatible dimensions.

{\color{black}The social cost functional is typically defined as (see \cite{wang2020indefinite})
\begin{equation*}
\tilde{J}_{\text{soc}}(\mathbf{u})= \sum_{j=1}^{N}\mathbb{E} \int_0^{\infty} \!\!\left(|x_j(\tau)-\Gamma x_{(N)}(\tau)|_Q^2+|u_j(\tau)|_{R}^2\right)\mathrm{d}\tau,
\end{equation*}
where $\mathbf{u}=(u_1,\cdots,u_N)$,} $x_{(N)}=(1/N)\sum_{j=1}^{N} x_j$ denotes the population average, $Q \geq 0$, $R > 0$, $\Gamma$ are weighting matrices.

In the presence of disturbances $v_i$ for each agent, the performance of $\tilde{J}_{\text{soc}}(\mathbf{u})$ may deteriorate. To reduce the sensitivity to these disturbances, an $L_2$-gain condition  
\begin{equation*} 
\textcolor{black}{\frac{\tilde{J}_{\text{soc}}(\mathbf{u})}{\sum_{j=1}^{N}\mathbb{E}\int_{0}^{\infty}|v_j(\tau)|^2\mathrm{d}\tau}\leq\gamma^2 }
\end{equation*}
is introduced by restricting $v_i(t)$ to the space of all $\mathcal{F}_t^i$-progressively measurable processes with values in $\mathbb{R}^{m_2}$ satisfying $\mathbb{E}\int_0^{\infty}|v_i(\tau)|^2\mathrm{d}\tau<\infty$, where $\mathcal{F}_t^i = \sigma(x_i(s),s\leq t)$, $t\geq0$. {\color{black}This condition ensures that the gain from the disturbance team vector $\mathbf{v}=[v_1^{\top},\cdots,v_N^{\top}]^{\top}$ to the evaluated signal $z=[z_1^{\top},\cdots,z_N^{\top}]^{\top}$ with each element $z_i=[(x_i-\Gamma x_{(N)})^{\top}Q^{\frac{1}{2}},
u_i^{\top}R^{\frac{1}{2}}]^{\top}$ is bounded by $\gamma$ in expectation, and is equivalently written as
\begin{equation}\label{eq:bound1}
{\tilde{J}_{\text{soc}}(\mathbf{u})}\leq\gamma^2 {\sum_{j=1}^{N}\mathbb{E}\int_{0}^{\infty}|v_j(\tau)|^2\mathrm{d}\tau}.
\end{equation}
In the classical robust control setting, the objective is to find the range of disturbance attenuation levels $\gamma>0$ for which there exists a control team $\mathbf{u}$ satisfying \eqref{eq:bound1} for all disturbance teams. Such $\gamma$ is called a viable attenuation level, and the smallest one is referred to as the minimum attenuation level.}

On the other hand, following the $H_{\infty}$ design fashion {\color{black}\cite{bacsar2008h,modares2014optimal}, the robust social optimization problem treats $\gamma>0$ as a predetermined parameter. The corresponding modified cost functional is then given by (see \cite{liang2022robust})
\begin{equation}\label{Jsoc}
J_{\text{soc}}(\mathbf{u},\mathbf{v})=\tilde{J}_{\text{soc}}(\mathbf{u})-\gamma^2 {\sum_{j=1}^{N}\mathbb{E}\int_{0}^{\infty}|v_j(\tau)|^2\mathrm{d}\tau}.
\end{equation}
A control team $\mathbf{u}$ satisfies \eqref{eq:bound1} for a given level $\gamma$ if and only if $J_{\text{soc}}(\mathbf{u},\mathbf{v})\leq 0$ for all  $\mathbf{v}$. Therefore, finding such $\mathbf{u}$ is equivalent to minimizing  \eqref{Jsoc} subject to $N$ agents’ dynamics \eqref{sys1}. Since this holds for all $\mathbf{v}$, we can consider the problem of finding $\mathbf{u}$ that minimizes the worst-case value of $J_{\text{soc}}(\mathbf{u},\mathbf{v})$.}

Accordingly, the robust social optimization problem can be formulated as a two-team zero-sum differential game \cite{liang2022robust}, where the control team $\mathbf{u}$ seeks to minimize the social cost, while the disturbance team $\mathbf{v}$ desires to maximize it. The solution corresponds to saddle point $(\mathbf{u}^*,\mathbf{v}^*)$ satisfying
\begin{equation}
J_{\text{soc}}(\mathbf{u}^*,\mathbf{v}^*) = \mathop{\inf_{\mathbf{u}}\sup_{\mathbf{v}}}J_{\text{soc}}(\mathbf{u},\mathbf{v}) = \sup_{\mathbf{v}}\inf_{\mathbf{u}}J_{\text{soc}}(\mathbf{u},\mathbf{v}).
\end{equation}

Depending on the different information structures available to the agents, their admissible strategies can be either centralized, denoted by $({\bf u},{\bf v})\in\mathcal{U}_{c}^u\times\mathcal{U}_{c}^v$, or decentralized, denoted by $({\bf u},{\bf v})\in\mathcal{U}_{d}^u\times\mathcal{U}_{d}^v$. The precise definitions of these admissible sets are standard and can be founded in \cite{wang2020indefinite,liang2022robust}. Within the set of admissible decentralized strategies, we introduce the concept of the mean field social optimum.
\begin{definition} 
A set of decentralized strategies $({\bf u}^{o},{\bf v}^{o}) \in \mathcal{U}_{d}^u \times \mathcal{U}_{d}^{v}$ is called a mean field social optimum if it has asymptotic robust social optimality, {\sl i.e., 
\begin{equation*}
\begin{aligned}
\left|\frac{1}{N}J_{\mathrm{soc}}({\bf u}^o,{\bf v}^o)-\frac{1}{N}\inf_{{\bf u}\in \mathcal{U}_{c}^u}\sup_{{\bf v}\in\mathcal{U}_{c}^v}J_{\mathrm{soc}}({\bf u},{\bf v})\right|=o(1).
\end{aligned}
\end{equation*}}
\end{definition}

Following the direct approach in \cite{liang2022robust}, the closed-loop decentralized strategies
\begin{numcases}{}
{u}^{o}_i(t)= -K_p^*x_i(t)-(K_s^*-\Upsilon^{-1}B^{\top}P^*)\bar{x}(t),\label{uo}\\
{v}^{o}_i(t)=L_p^*x_i(t)+(L_s^*-L_p^*)\bar{x}(t),~~1\leq i\leq N,\label{do}
\end{numcases}
together with the closed-loop mean field dynamics
\begin{equation}\label{xbar}
\dot{\bar{x}}(t) = \left[ A-B (K_s^* + \Upsilon^{-1} D^{\top} P^*C)+GL_s^*\right]\bar{x}(t),
\end{equation}
$\bar{x}(0)=\bar{x}_0$, constitute a mean field social optimum. Here, the gain matrices {\color{black}$K_p^*$, $K_s^*$, $L_p^*$}, and {\color{black}$L_s^*$} are defined as 
% \begin{numcases}{}
% &K_p^* \triangleq \Upsilon^{-1}(B^{\top}P^*+D^{\top}P^*C),\label{gain_op1}\\
% &L_p^* \triangleq  \frac{1}{\gamma^2}G^{\top}P^*,\label{gain_op2}\\
% &K_s^* \triangleq  \Upsilon^{-1}B^{\top}S^*,\label{gain_op3}\\
% &{L}_s^* \triangleq  \frac{1}{\gamma^2}G^{\top}S^*,\label{gain_op4}
% \end{numcases}
\begin{equation*} 
\left\{\begin{aligned}
&{\color{black}K_p^*} \triangleq \Upsilon^{-1}(B^{\top}P^*+D^{\top}P^*C),~~{\color{black}K_s^*} \triangleq  \Upsilon^{-1}B^{\top}S^*,\\
&{\color{black}L_p^*} \triangleq  \frac{1}{\gamma^2}G^{\top}P^*,~~{\color{black}{L}_s^*}\triangleq  \frac{1}{\gamma^2}G^{\top}S^*,
\end{aligned}\right.
\end{equation*}
with $\Upsilon\triangleq R+D^{\top}{\color{black}P^*}D>0$. $P^*$ and $S^*$ are the unique positive semidefinite stabilizing solutions to the following equations
\begin{equation}\label{are1}
\begin{aligned}
\!\!\!\!\!&A^{\top}\!{\color{black}P^*}\!+\!{\color{black}P^*}\!A\!+\!C^{\top}\!{\color{black}P^*}C\!+\!Q\!+\!\frac{1}{\gamma^2} {\color{black}P^*}GG^{\top}\!{\color{black}P^*}~~~~~~~~~~~~~\\
&~~~~~-\!({\color{black}P^*}\!B\!+\!C^{\top}\!{\color{black}P^*}D)\Upsilon^{-1}(B^{\!\top}{\color{black}P^*}+D^{\top}{\color{black}P^*}C)=0,
\end{aligned}
\end{equation}
\begin{equation}\label{are3}
\begin{aligned}
\!\!\!\!\!A_s^{\top}\!{\color{black}S^*}\!\!+ \!{\color{black}S^*}A_s\!+\!Q_s\!+\!\frac{1}{\gamma^2}{\color{black}S^*}GG^{\top}{\color{black}S^*}\!\!-\!{\color{black}S^*}B\Upsilon^{-1}B^{\top}\!{\color{black}S^*}\!=\!0,
\end{aligned}
\end{equation}
where $A_s=A-B\Upsilon^{-1}D^{\top}P^*C$, $Q_{\Gamma}=-\Gamma^{\top}Q\Gamma+\Gamma^{\top}Q+Q\Gamma$ and $Q_s=Q-Q_{\Gamma}+(C-D\Upsilon^{-1}D^{\top}P^*C)^{\top}P^*(C-D\Upsilon^{-1}D^{\top}P^*C)+C^{\top}P^*D\Upsilon^{-1}R\Upsilon^{-1}D^{\top}P^*C\geq0$.  

\textcolor{black}{Note that the quadratic terms involving the perturbation channel make the Riccati operators indefinite, which  complicates the analysis and computation of stabilizing solutions. Moreover, in practical applications, the exact parameters of the agents’ dynamics are often unknown or only partially known. Solving such Riccati-type equations, particularly under model uncertainty, poses significant challenges. Therefore, our main goal is to develop an iterative method that does not require full knowledge of the system dynamics while ensuring convergence and robustness.}  

% especially in presence of disturbances and limited prior knowledge of the agents' dynamics ({\sl i.e., matrices $A,B,G,C,D$}), which becomes a bottleneck in practical applications. To overcome this challenge, our main goal is to develop an iterative method that does not rely on an exact dynamical model and has guaranteed convergence and robustness.
  
\section{A unified iterative framework for solving indefinite Riccati-type equations}\label{sec:main result1}
In this section, we present a unified dual-loop iterative framework for solving both the indefinite SARE and the indefinite ARE. In the outer-loop iteration, a sequence of $H_2$-type (S)AREs is introduced, whose solutions converge to the stabilizing solution of the target indefinite (S)ARE. In the inner-loop iteration, each outer-loop solution is further approximated by solving a sequence of linear matrix equations. 

\subsection{Useful concepts and results}
{\color{black}To facilitate the development, we define the generalized Lyapunov operator $\mathscr{L}_{[{K},{L},\mathcal{S}_p]}:\mathbb{S}^n\mapsto\mathbb{S}^n$ for given gains $K,L$ and system \eqref{sys1} (abbreviated as $\mathcal{S}_p\triangleq [A,B,G|C,D]$) by  
\begin{equation}\label{opt:Lyapunov}
 \begin{aligned}
\!\!\!\!\!\!\!\mathscr{L}_{[{K},{L},\mathcal{S}_p]}(P)\!\triangleq&({A}\!-\!{B}{K}\!+\!{G}{L})^{\top}P\!+\!P({A}\!-\!{B}{K}\!+\!{G}{L})\\
\!\!\!\!\!\!\!&+({C}-{D}{K})^{\top}P({C}-{D}{K}) 
\end{aligned} 
\end{equation}
with spectrum
\begin{equation*}
\sigma(\mathscr{L}_{[{K},{L},{S}_p]})\!=\!\{\lambda\in\mathbb{C}:\!\mathscr{L}_{[{K},{L},{S}_p]}(P)\!=\!\lambda P, P\!\in\!\mathbb{S}^n,P\!\neq\!0\}.
\end{equation*}
The adjoint operator $\mathscr{L}_{[K,L,\mathcal{S}_p]}^*$ is given by
\begin{equation*}
 \begin{aligned}
\!\!\!\!\!\!\!\mathscr{L}_{[{K},{L},\mathcal{S}_p]}^*(P)\!=&({A}\!-\!{B}{K}\!+\!{G}{L})P\!+\!P({A}\!-\!{B}{K}\!+\!{G}{L})^{\top}\\
\!\!\!\!\!\!\!&+({C}-{D}{K})P({C}-{D}{K})^{\top},
\end{aligned}
\end{equation*}
which is commonly used to analyze asymptotic mean
square stability of It\^{o} differential systems via its spectrum \cite{zhang2004stabilizability}. Note that all the coefficient matrices considered here are real, therefore $\sigma(\mathscr{L}_{[K,L,\mathcal{S}_p]})$ is identical to $\sigma(\mathscr{L}^*_{[K,L,\mathcal{S}_p]})$.  

Let $\mathcal{U}_p \triangleq \{B, R, C,D\}$ and $\mathcal{V} \triangleq \{G,\gamma^2I\}$, and define the gain mappings by
\begin{align}
&\mathscr{K}^1_{\mathcal{U}_p}(P) \triangleq({R}+{D}^{\top}P{D})^{-1}\left({B}^{\top}P+{D}^{\top}P{C}\right),\\
&\mathscr{K}^2_{\mathcal{V}}(P)\triangleq\frac{1}{\gamma^2} {G}^{\top} P.
\end{align}
With these, the indefinite SARE (\ref{are1}) can be rewritten as
\begin{subequations}{}\label{are_operator}
\begin{empheq}[left={\left\{\begin{aligned}},right={\end{aligned}\right.}]{align}
&\mathscr{L}_{[{K}_p^*,{L}_p^*,\mathcal{S}_p]}(P^*)+Q-\gamma^2|L_p^*|^2+|K_p^*|_{R}^{2}=0,\label{are_operator1}\\
&{K}_p^*=\mathscr{K}_{\mathcal{U}_p}^1(P^*),\label{are_operator3}\\
&{L}_p^* =\mathscr{K}_{\mathcal{V}}^2(P^*).\label{are_operator2}
\end{empheq}
\end{subequations}
Next, replacing $L^*$ in \eqref{are_operator} with a constant matrix ${L}$ yields the following $H_2$-type SARE
\begin{subequations}\label{general:oueter-loop}
\begin{empheq}[left={\left\{\begin{aligned}},right={\end{aligned}\right.}]{align}
&\mathscr{L}_{[K_{{L}},{L},\mathcal{S}_p]}(P_{{L}})+ {Q} -\gamma^2|{L}|^2+|{K}_{{L}}|_{R}^2=0,\label{general:oueter-loop1}\\
&{K}_{{L}} = \mathscr{K}_{\mathcal{U}_p}^1(P_{{L}}). \label{general:oueter-loop2}
\end{empheq}
\end{subequations}
Based on this equation, we further fix ${K}_{{L}}$ to a constant matrix ${K}$ to obtain the Lyapunov-type equation
\begin{equation}\label{general:inner-loop}
\mathscr{L}_{[{K},{L},\mathcal{S}]}(P_{{K}})+ {Q} -\gamma^2|{L}|^2+|{K}|_{R}^2=0.
\end{equation}}
 \begin{figure}[H]
\centering
\includegraphics[scale=0.32]{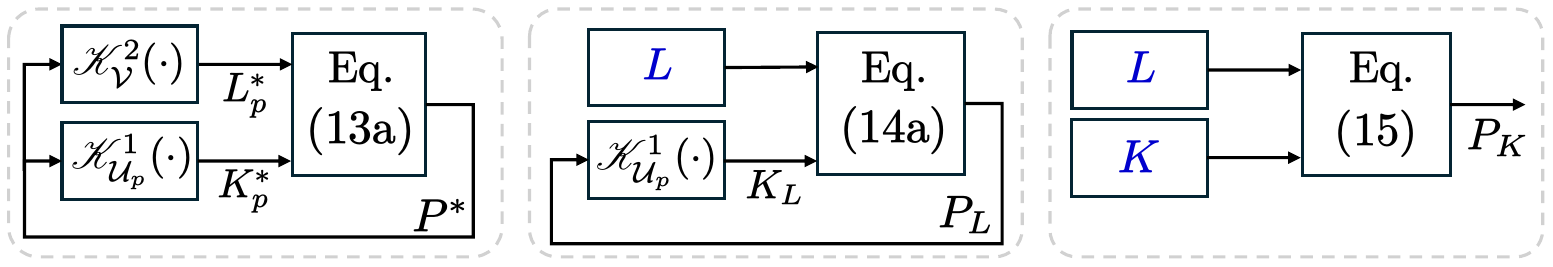}
\centering
\caption{Structures of different equations: (left) Indefinite SARE; (middle) $H_2$-type SARE; (right) Lyapunov-type equation.}\label{f:digram1}
\end{figure}
{\color{black}As illustrated in Fig. \ref{f:digram1}, the transformation from  (\ref{are_operator}) $\to$ \eqref{general:oueter-loop} $\to$ (\ref{general:inner-loop}) opens the closed loops in sequence. The resulting two equations provide the basis for the subsequent iterative procedures by introducing new loops with specific logic. 

Furthermore, we need the following standard assumption.
\begin{assume}\label{assume:3-1}
$[A,B,0|C,D]$ ( or simply $[A,B|C,D]$), is stabilizable and $[{A},{Q}|{C}]$ is exactly detectable.
\end{assume}
\begin{remark}
By the spectrum method, necessary and sufficient conditions for stochastic stablizability (in the mean square sense) and exact detectability are given in \cite[Theorem 1]{zhang2004stabilizability} and \cite[Theorem 3.1]{zhang2008generalized}, respectively. Note that the exact detectability is a weaker concept than stochastic detectability, the latter being defined as the dual of stabilizability (see \cite[Definition 3]{chen2004stochastic}). More specifically, from \cite[Proposition 1]{zhang2005spectral} and \cite[Theorem 3.1]{zhang2008generalized}, it follows that stochastic detectability implies exact detectability, but the converse does not hold. These properties are fundamental for analyzing the existence and uniqueness of stabilizing solutions to (S)AREs and generalized Lyapunov equations with (semi)definite weighting matrices (see \cite[Theorems 3.2 \& 4.1]{zhang2008generalized}) and will be used in subsequent analyses. For clarity, we restate the relevant consequences below using the operator and notation introduced above.
\end{remark}
\begin{proposition}\label{propo:stabilization}
$[A,B|C,D]$ is stabilizable iff there exists $K\in\mathbb{R}^{m_1\times n}$ such that the spectrum of $\mathscr{L}_{[K,0,\mathcal{S}_p]}$ lies entirely in the open left half-plane $\mathbb{C}^-$.
\end{proposition} 
\begin{proposition} \label{propo:detectable_sufficient}
$[A,Q|C]$ is exactly detectable iff there does not exist nonzero $P\in\mathbb{S}^n$ such that 
\begin{equation*}
\mathscr{L}_{[{0},{0},\mathcal{S}_p]}(P)=\lambda P,~~QP=0,~~\mathrm{Re}(\lambda)\geq0.
\end{equation*}
\end{proposition}
\begin{proposition}\label{propo:SARE}
If $[A,B|C,D]$ is stabilizable and $[A,Q|C]$ is exactly detectable, then the following SARE
\begin{equation*}  
\left\{\begin{aligned}
&\mathscr{L}_{[K,{0},\mathcal{S}_p]}(P)+Q+|K|_R^2=0,~~Q\geq0,~~R>0,\\
&K = \mathscr{K}_{\mathcal{U}_p}^1(P) 
\end{aligned}\right.
\end{equation*}
\end{proposition}
has a unique solution $P\geq0$ such that $\sigma(\mathscr{L}_{[K,{0},\mathcal{S}_p]})\subset\mathbb{C}^-$.
\begin{proposition}\label{propo:GLE}
If $[A,Q|C]$ is exactly detectable and the generalized Lyapunov equation
\begin{equation*} 
\mathscr{L}_{[{0},{0},\mathcal{S}_p]}(P)+ Q=0,~~Q\geq0
\end{equation*}
has a solution $P\geq0$, then $\sigma(\mathscr{L}_{[{0},{0},\mathcal{S}_p]})\subset\mathbb{C}^-$.
\end{proposition}

For (\ref{general:oueter-loop}), we define the admissible set of ${L} $ as 
\begin{align*}
\mathcal{W}\triangleq\{&{L}\in\mathbb{R}^{m_2\times n}|\text{Eq. \eqref{general:oueter-loop} has a unique  solution}\\
&\text{ such that $0 \leq P_{{L}}\leq P^*$ and $\sigma(\mathscr{L}_{[{K}_{{L}},{L},\mathcal{S}_p]}) \subset \mathbb{C}^-$}\}. \end{align*}
Let $L'=\mathscr{K}_{\mathcal{V}}^2(P_{{L}})$, we then conclude the following result, with proof in Appendix \ref{pflem:used in Theorem1}.
\begin{lemma}\label{lem:used in Theorem1}
If ${L}\in\mathcal{W}$, then $\sigma(\mathscr{L}_{[K_p^*,L',\mathcal{S}_p]})\subset\mathbb{C}^-$.
\end{lemma}
This result will be used in the subsequent outer-loop iteration convergence analysis. Moreover, for \eqref{general:inner-loop}, we define the admissible set of ${K}$ for a given ${L}\in\mathcal{W}$ as
\begin{equation*}
\begin{aligned}
\mathcal{Z}({L})\triangleq \{&{K}\in\mathbb{R}^{m_1\times n}| \sigma(\mathscr{L}_{[{K},{L},\mathcal{S}_p]})\subset\mathbb{C}^{-}\}.
\end{aligned}
\end{equation*}}
  
\subsection{Introduction of outer-loop iteration}
To approximate the stabilizing solution $P^*\geq0$ of Eq. \eqref{are1}, we design an iterative scheme based on Eq. \eqref{general:oueter-loop}. In the k-th iteration, the constant matrix ${L}$ is replaced by the (k-1)-th approximation ${L}_p^{k-1}$ of $L_p^*$, leading to the following equation
\begin{subequations}{}\label{outer_eq}
\begin{empheq}[left={\left\{\begin{aligned}},right={\end{aligned}\right.}]{align}
&{\mathscr{L}}_{[{K}^{k}_p,{L}_p^{k-1},\mathcal{S}_p]}(P^{k}) +Q_p^{k-1}+ |{K}_p^{k}|_{R}^2=0,\label{outer_eq:xik+1}\\
&K_p^{k}=\mathscr{K}_{\mathcal{U}_p}^1(P^{k}),\label{outer_eq:kik+1}
\end{empheq}
\end{subequations}
where $Q_p^k \triangleq Q-\gamma^2|L_p^k|^2$ may be indefinite, recursively,
\begin{equation}\label{outer_eq2}
{L}_p^{k-1}=\mathscr{K}_{\mathcal{V}}^2(P^{k-1}),~~k=1,2,\cdots.
\end{equation}
\begin{figure}[!htb]
\centering
\includegraphics[scale=0.31]{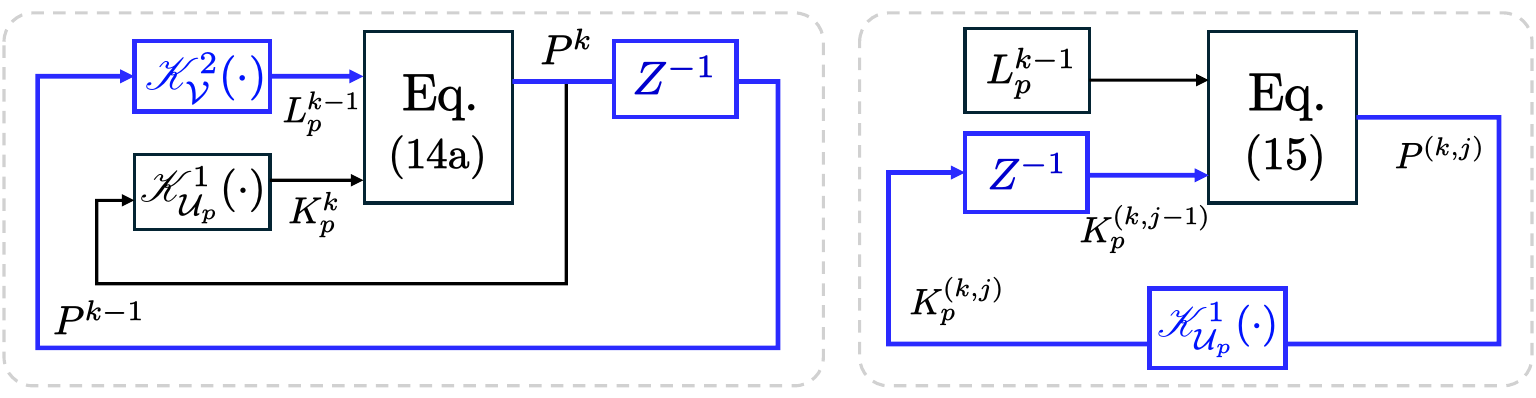}
\centering
\caption{Schematic diagram of the iterative schemes: (left) outer-loop iteration; (right) inner-loop iteration.}\label{f:diag2}
\end{figure}
As shown in Fig. \ref{f:diag2} (left), this outer-loop iteration introduces a new closed-loop structure with a one-step delay compared to \eqref{general:oueter-loop}. The process generates the sequences $\{P^k,L_p^k,K_p^k\}_{k=1}^{\infty}$. Before studying their convergence to $(P^*,L^*_p,K^*_p)$, we first establish an important intermediate result.
\begin{lemma}\label{lemma:W}
If $L_p^{k-1}\in\mathcal{W}$, then ${L}_p^k\in\mathcal{W}$ and $P^{k}\leq P^{k+1}$.
\end{lemma}

\bproof{\color{black}
For brevity, define $R_p^k \triangleq R+D^{\top}P^kD$,
\begin{equation*}
\begin{aligned}
&\tilde{P}^{k} \triangleq P^{k+1}-P^{k}, ~\tilde{K}_p^{k} \triangleq K_p^{k+1}-K_p^{k}, ~\tilde{L}_{p}^{k} \triangleq L_{p}^{k+1}-L_p^{k},\\
&\Delta P^k \triangleq P^*-P^k, ~\Delta K_p^k \triangleq K_p^*-K_p^k, ~\Delta L_p^k \triangleq L_p^*-L_p^k.
\end{aligned}
\end{equation*}
To complete the proof, we first transform Eq. \eqref{outer_eq} with the indefinite weighting matrix $Q_p^{k-1}$ for $P^k$ into a SARE with a definite state weighting matrix for $\tilde{P}^{k}$. Then, by using Proposition \ref{propo:SARE}, we establish that $\tilde{P}_k\geq0$ and $\sigma(\mathscr{L}_{[K_p^{k+1},L_{p}^k,\mathcal{S}_p]})\subset\mathbb{C}^-$. Finally, we deduce   $P^{k+1}\leq P^*$.

We begin by applying the operator $\mathscr{L}_{[{K}_p^{k},{L}_p^{k},\mathcal{S}_p]}$ and completing the square in (\ref{outer_eq:xik+1}) with $k$ replaced by $k+1$, yielding
\begin{equation}\label{th1eq4}
\begin{aligned}
{\mathscr{L}}_{[{K}_p^{k},{L}_p^{k},\mathcal{S}_p]}(P^{k+1}) +{Q}_p^{k}-|{\tilde{{K}}_p^{k}}|^{2}_{{R}_p^{k+1}} +|{K}_p^{k}|_{{R}}^2=0.
\end{aligned}
\end{equation}
Similarly, for index $k$, Eq. \eqref{outer_eq:xik+1} reads
\begin{equation*} 
\begin{aligned}
{\mathscr{L}}_{[{K}_p^{k},{L}_p^{k},\mathcal{S}_p]}(P^{k}) + {Q}_p^{k} -\gamma^2|\tilde{{L}}_p^{k-1}|^2+|{K}_p^{k}|_{{R}}^2=0.
\end{aligned}
\end{equation*}
Subtracting these equations eliminates $Q^k_p$ and results in
\begin{equation}\label{th1eq6}
\begin{aligned}
&{\mathscr{L}}_{[{K}_p^{k},{L}_p^{k},\mathcal{S}_p]}(\tilde{P}^{k})+\tilde{Q}^{k-1}_p-|\tilde{{K}}_p^{k}|^2_{{R}_p^{k+1}}=0,
\end{aligned}
\end{equation} 
where $\tilde{Q}^{k-1}_p=\gamma^2|\tilde{{L}}_p^{k-1}|^2\geq0$. 

To facilitate the equation transformation, we establish a key identity linking $\tilde{K}_p^k$ and $\tilde{P}^k$. To proceed, rewrite Eq. \eqref{outer_eq:kik+1} by
\begin{equation*}
\begin{aligned}
\tilde{K}_p^k&=(R_p^{k+1})^{-1}(B^{\top}P^{k+1}+D^{\top}P^{k+1}(C-DK^{k}_p))\\
&~~~~+(R_p^{k+1})^{-1}D^{\top}P^{k+1}DK^{k}_p-K_p^k.
\end{aligned}
\end{equation*}
This can be simplified as
\begin{equation*}
\tilde{K}_p^k= (R_p^{k+1})^{-1}(B^{\top}P^{k+1}+D^{\top}P^{k+1}(C-DK^{k}_p)-RK_p^k).
% &=(R+D^{\top}P^{k+1}D)^{-1}(B^{\top}\tilde{P}^k+D^{\top}\tilde{P}^k(C-DK^{k}_p)).
\end{equation*}
Let  ${C}_p^k \triangleq {C}-{D}{K}_p^k$, and recall that
\begin{equation*}
RK_p^k = (B^{\top}P^{k}+D^{\top}P^k {C}_p^k), ~R_p^{k+1}=R_p^{k}+D^{\top}\tilde{P}^kD.
\end{equation*}
Hence, we arrive at the pivotal formula
\begin{align}
\tilde{{K}}_p^{k} =(R_p^k+D^{\top}\tilde{P}^kD)^{-1}(B^{\top}\tilde{P}^k+D^{\top}\tilde{P}^kC_p^k).\label{ktilde}
\end{align}
Substituting Eq. (\ref{ktilde}) into the operator $\mathscr{L}_{[{K}_p^k,{L}_p^k,\mathcal{S}_p]}(\tilde{P}^k)$ gives
\begin{equation*}
\begin{aligned}
&\mathscr{L}_{[{K}_p^{k},{L}_p^{k},\mathcal{S}_p]}(\tilde{P}^{k})=(A_k-B\tilde{K}_p^k)^{\top}\tilde{P}^k+\tilde{P}^k(A_k-B\tilde{K}_p^k)\\
&~~ + (C_p^k \!-\! D^{\top} \tilde{K}_p^k)^{\top} \tilde{P}^k(C_p^k \!-\! D^{\top}\tilde{K}_p^k) + |\tilde{{K}}_p^{k}|_{{R}_p^{k+1}}^2 \!+ |\tilde{{K}}_p^{k}|_{{R}_p^{k}}^2,
\end{aligned} 
\end{equation*}
where $A_k = A -B K_p^k + G L_p^k$. Therefore, Eq. \eqref{th1eq6} can now be written as the standard SARE
\begin{equation}\label{th1eq7}
\begin{aligned}
\!\!\!&A_k^{\top}\!\tilde{P}^k\!+\!\tilde{P}^k\!A_k+(C_p^k)^{\top}\!\tilde{P}^k C_p^k\!+ \tilde{Q}^{k-1}_p\!-\!(\tilde{P}^{k}\!B\!\\
\!\!\!&+\!(C_p^k)^{\top}\!\tilde{P}^k\!D)(R_p^k\!+\!D^{\top}\!\tilde{P}^k\!D)^{-1}(B^{\top}\!\tilde{P}^k\!\!+\!\!D^{\top}\!\tilde{P}^kC_p^k)\!=\!0.
\end{aligned}
\end{equation}
Since $\tilde{Q}_p^{k-1}\geq0$, Proposition \ref{propo:SARE} ensures that \eqref{th1eq7} has a unique positive semidefinite stabilizing solution if (\romannumeral1) $[A_k,B|C_p^k,D]$ is stabilizable; and (\romannumeral2) $[A_k,\tilde{Q}_p^k|C_p^k]$ is exactly detectable.

Starting with $K=\Delta {K}_{p}^{k}$, we have
\begin{equation*}
\begin{aligned}
&(A_k-BK)^{\top}P+P(A_k-BK)\\
&+(C_p^k-DK)^{\top}P(C_p^k-DK)=\mathscr{L}_{[{K}_p^*,{L}_p^{k},\mathcal{S}_p]}(P).
\end{aligned}
\end{equation*}
Since ${L}_{p}^{k-1}\in \mathcal{W}$, Lemma \ref{lem:used in Theorem1} ensures that $\sigma(\mathscr{L}_{[{K}_p^*,{L}_p^{k},\mathcal{S}_p]}) \subset \mathbb{C}^-$. Therefore, by Proposition \ref{propo:stabilization}, condition (\romannumeral1) is verified.
 
For condition (\romannumeral2), suppose by contradiction that exact detectability does not hold. Then there exists $P\!\neq\!0$ such that
\begin{equation*}
A_k^{\top}P+PA_k+(C_p^k)^{\top}PC_p^k=\lambda P, \quad \mathrm{Re}(\lambda)\geq0,
\end{equation*}
and $\tilde{Q}_p^{k-1}P=0$, which yields $L_p^kP=L_{p}^{k-1}P$. Consequently,
\begin{equation*}
\mathscr{L}_{[K_p^k,L_p^{k-1},\mathcal{S}_p]}^*(P) = \lambda P,~~\mathrm{Re}(\lambda)\geq0,
\end{equation*}
which contradicts $\sigma(\mathscr{L}_{[K_p^k,L_p^{k-1},\mathcal{S}_p]})\subset\mathbb{C}^-$. Therefore, condition (\romannumeral2) holds. 

Since both conditions are satisfied, we can conclude that $\tilde{P}^{k}\geq 0\Rightarrow {P}_p^{k+1}\geq{P}_p^{k}\geq0$ and $\sigma(\mathscr{L}_{[{K}_p^{k+1},{L}_p^{k},\mathcal{S}_p]})\subset \mathbb{C}^{-}$.
 
To establish the upper bound ${P}_p^{k+1}\leq{P}^*$, we subtract (\ref{th1eq4}) from (\ref{are_operator1}), which yields
\begin{equation*} 
\begin{aligned}
\mathscr{L}_{[{K}_p^*,{L}_p^{k},\mathcal{S}_p]}(\Delta{P}_p^{k+1}) +\gamma^2|\Delta{L}_p^{k}|^2+|\Delta{K}_p^{k+1}|_{{R}_p^{k}}^2=0.
\end{aligned}
\end{equation*} 
Since ${L}_p^{k-1}\in\mathcal{W}$ ensures ${R}_p^{k}>0$ and $\sigma(\mathscr{L}_{[{K}_p^*,{L}_p^{k},\mathcal{S}_p]})\subset\mathbb{C}^-$ (by Lemma \ref{lem:used in Theorem1}), it follows from \cite[Theorem 3.2.3]{sun2020stochastic} that $\Delta {P}^{k+1}\geq0$, {\sl i.e., ${P}^{k+1}\leq{P}^*$}.  }
\eproof

\begin{theorem}\label{thm:convergence-outer-loop}
Suppose Assumption \ref{assume:3-1} holds. {\color{black}Let $P^*\geq0$ be the unique stabilizing solution to Eq. \eqref{are1}, and let $K_p^*$ and $L_p^*$ be determined by \eqref{are_operator3} and \eqref{are_operator2}, respectively.} Let the sequences $\{P^k,{K}_p^{k},{L}_p^{k}\}_{k=1}^{\infty}$ be generated by recursively solving (\ref{outer_eq}) and (\ref{outer_eq2}) with the initial conditions $P^0=0$ and $L_p^0\!=\!\mathscr{K}_{\mathcal{V}}^2(P^0)\!=\!0$. Then, $\lim\limits_{k\rightarrow\infty}{P}^{k}=P^*$, $\lim\limits_{k\rightarrow\infty}{K}_p^{k}={K}_p^*$, and $\lim\limits_{k\rightarrow\infty}{L}_p^{k}={L}_p^*$.
\end{theorem}
\bproof
We first show, by mathematical induction, that $L_p^{k-1}\in\mathcal{W}$ and $P^{k-1}\leq P^{k}$ hold for all $k\in\mathbb{N}_+$.
\begin{itemize}
{\color{black}\item[i).] For $k=1$, Eq. \eqref{outer_eq} at $k=1$ reduces to 
\begin{equation}\label{th1eq1}
\begin{aligned}
\!\!\!\!\!\mathscr{L}_{[{K}_p^1,0,\mathcal{S}_p]}(P^1)\!+\!{Q}\!+\!|{K}_p^1|_{{R}}^2\!=\!0,~{K}_p^{1}\!=\!\mathscr{K}_{\mathcal{U}_p}^1(P^1).
\end{aligned}
\end{equation}
Under Assumption \ref{assume:3-1}, this equation has a solution $P^1\geq 0=P^0$ such that $\sigma(\mathscr{L}_{[{K}_p^1,0,\mathcal{S}_p]})\subset\mathbb{C}^{-}$.

To establish $P^1\leq P^*$, we recall that $P^*$ satisfies
\begin{equation}\label{th1eq1-3}
\mathscr{L}_{[{K}_p^*,0,\mathcal{S}_p]}(P^*)+Q+\gamma^2|L_p^*|^2+|K_p^*|_R^2=0.
\end{equation}
Subtracting Eq. (\ref{th1eq1}) from Eq. (\ref{th1eq1-3}) yields 
\begin{equation}\label{th1eq2}
\begin{aligned}
\mathscr{L}_{[{K}_p^*,0,\mathcal{S}_p]}(\Delta P^1)+ \gamma^2|{L}_p^*|^2+|\Delta{K}_p^1|^2_{{R}_p^1}=0.
\end{aligned}
\end{equation}
Since $\sigma(\mathscr{L}_{[K_p^*,0,\mathcal{S}_p]})\subset\mathbb{C}^-$ can be easily verified and $R>0$, it follows that $\Delta P^1\geq0$. Thus, $L_p^0\in\mathcal{W}$ and $P^0\leq P^1$. Applying Lemma \ref{lemma:W} yields $L_p^1 \in \mathcal{W}$ and $P^1 \leq P^2$.}\\
\item[ii).] Assume for some $q>1$, $L_p^{k-1}\in\mathcal{W}$ and $P^{k-1}\leq P^{k}$ hold for $k = q$. Then Lemma \ref{lemma:W} ensures $L_p^{q}\in\mathcal{W}$ and $P^{q}\leq P^{q+1}$. 
\end{itemize}

{\color{black}By induction, $L_p^{k-1}\in\mathcal{W}$ and $P^{k-1}\leq P^{k}$ hold for all $k\in\mathbb{N}_+$, so $\{P^k\}$ is monotonically increasing and bounded above by $P^*$. Therefore, the limit exists, {\sl i.e.}, there exists $P^{\infty}$ such that $\lim_{k\rightarrow\infty}P^{k}=P^{\infty}$. Taking the limit in \eqref{outer_eq} gives
\begin{equation*} 
\begin{aligned}
\!\!\!\!\!&A^{\top}\!P^{\infty}\!+\!P^{\infty}\!A\!+\!C^{\top}\!P^{\infty}C\!+\!Q\!+\!\frac{1}{\gamma^2}\!P^{\infty}GG^{\top}\!P^{\infty}\!-\!(P^{\infty}\!B\\
&+\!C^{\top}\!P^{\infty}D)(R+D^{\top}P^{\infty}D)^{-1}\!(B^{\!\top}P^{\infty}+D^{\top}P^{\infty}C)=0.
\end{aligned}
\end{equation*}
Since $P^*$ is the unique positive semidefinite solution of this equation, we have $P^{\infty}=P^*$. Hence, it follows from \eqref{outer_eq:kik+1} and \eqref{outer_eq2} that $\lim_{k\rightarrow\infty}{K}_p^{k}\!=\!{K}_p^*$ and $\lim_{k\rightarrow\infty}{L}_p^{k}\!=\!{L}_p^*$.}
\eproof
\begin{remark}
To the best of our knowledge, \cite{dragan2011computation} was the first to provide a convergent iterative scheme for the indefinite SARE by constructing a sequence of SAREs with definite quadratic terms, which extended the deterministic result in \cite{lanzon2008computing}. Our proposed sequence of SAREs, however, is fundamentally different and provably not equivalent to the method in \cite{dragan2011computation}. In particular, if the method of \cite{dragan2011computation} were applied to solve \eqref{are1}, the control gain update law \cite[Eq. (12)]{dragan2011computation} would take the form
\begin{equation*}
{K}_p^k = (I+({R}_p^k)^{-1}{D}^{\top}P^k{D})^{-1}({R}_p^{k})^{-1}({B}^{\top}P^k+{D}^{\top}P^k{C}),
\end{equation*}
which is distinctly different from our update law (\ref{outer_eq:kik+1}). Moreover, our analysis is built on exact detectability, which is weaker than the stochastic detectability condition used in \cite{dragan2011computation}. Finally, our iterative framework not only establishes convergence but also provides a foundation for developing a date-driven algorithm to solve the indefinite SARE.  
\end{remark}

 {\color{black}Let $\mathcal{S}_s\triangleq[A_s,B|0,0]$ (or simply $(A_s,B)$) and $\mathcal{U}_s\triangleq\{B,\Upsilon,0,0\}$, the indefinite ARE \eqref{are3} can be rewritten as
 \begin{subequations}{}\label{are2_operator}
\begin{empheq}[left={\left\{\begin{aligned}},right={\end{aligned}\right.}]{align}
&\mathscr{L}_{[{K}_s^*,{L}_s^*,\mathcal{S}_s]}(S^*)+Q_s-\gamma^2|L_s^*|^2+|K_s^*|_{\Upsilon}^{2}=0,\label{are2_operator1}\\
&{K}_s^*=\!\mathscr{K}_{\mathcal{U}_s}^1(S^*),\label{are2_operator3}\\
&{L}_s^* =\mathscr{K}_{\mathcal{V}}^2(S^*).\label{are2_operator2}
\end{empheq}
\end{subequations}
Since there exists $K=\Upsilon^{-1}B^{\top}P^*$ such that $A_s-BK$ is Hurwitz, we have that $\mathcal{S}_s$ is stabilizable. The deterministic outer-loop iteration for approximating the unique stabilizing solution $S^*\geq0$ of \eqref{are3} is then given by
\begin{subequations}{}\label{outer_eq:ARE}
\begin{empheq}[left={\left\{\begin{aligned}},right={\end{aligned}\right.}]{align}
&{\mathscr{L}}_{[{K}^{k}_s,{L}_s^{k-1},\mathcal{S}_s]}(S^{k})\!+\!{Q}_s\!-\!\gamma^2|{L}_s^{k-1}|^{2}\!+\!|{K}_s^{k}|_{\Upsilon}^2\!=\!0,\label{outer_eq:are_xik+1}\\
&K_s^{k}=\mathscr{K}_{\mathcal{U}_s}^1(S^{k}),\label{outer_eq:are_kik+1}
\end{empheq}
\end{subequations}
together with the recursion
\begin{equation}\label{outer_eq2:are}
~~{L}_s^{k-1}=\mathscr{K}_{\mathcal{V}}^2(S^{k-1}),~~k=1,2,\cdots.
\end{equation}
We impose the following assumption. 
\begin{assume}\label{assume:are}
$(A,Q^{\frac{1}{2}}(I-\Gamma))$ is detectable.
\end{assume}
\begin{remark}
 According to the equivalent statements of stabilizability and detectability in the deterministic case \cite{zhou1996robust}, the conclusions of Propositions \ref{propo:stabilization}-\ref{propo:detectable_sufficient} can be specialized as follows: 1). $(A_s,B)$ is stabilizable iff there exists $K\in\mathbb{R}^{m_1\times n}$ such that $\sigma(\mathscr{L}_{[K,0,\mathcal{S}_s]})\subset\mathbb{C}^-$; 2). $(A_s,Q)$ is detectable iff there does not  exist a nonzero $P\in\mathbb{S}^n$ satisfying $\mathscr{L}_{[0,0,\mathcal{S}_s]}(P)=\lambda P$, $QP=0$, $\mathrm{Re}(\lambda)\geq0$.
By invoking these results together with the classical deterministic LQR result \cite[Theorem 3]{kucera1972contribution}, the corresponding arguments in the proof of Theorem \ref{thm:convergence-outer-loop} carry over directly to the deterministic case. This immediately yields the following deterministic counterpart of Theorem \ref{thm:convergence-outer-loop}.
\end{remark}
\begin{corollary}\label{coro:1}
Suppose the conditions of Theorem \ref{thm:convergence-outer-loop} and Assumption \ref{assume:are} hold. Let $S^*\geq0$ be the unique solution to \eqref{are3}, and let  $K_s^*$ and $L_s^*$ be determined by \eqref{are2_operator3} and \eqref{are2_operator2}, respectively. Let $\{S^k,L_s^k,K_s^k\}_{k=1}^{\infty}$ be generated by recursively solving \eqref{outer_eq:ARE} and \eqref{outer_eq2:are} with $S^0=0$ and $L_s^0=\mathscr{K}_{\mathcal{V}}^2(S^0)=0$. Then,  $\lim\nolimits_{k\rightarrow\infty}S^k\!=\!S^*$, $\lim\nolimits_{k\rightarrow\infty}K_s^k\!=\!K_s^*$, and $\lim\nolimits_{k\rightarrow\infty}L_s^k\!=\!L_s^*$.
\end{corollary}}
 
\subsection{Introduction of inner-loop iteration}
To effectively solve the stabilizing solution $P^k$ of Eq. (\ref{outer_eq}), we introduce an inner-loop iteration based on (\ref{general:inner-loop}), as illustrated in Fig. \ref{f:diag2} (right). During the $(k,j)$-th inner-loop iteration, ${L}$ is fixed by ${L}_p^{k-1}$, while ${K}$ is replaced by the $(k,j-1)$-th approximation of $K_p^k$. As a result, the $(k,j)$-th approximation of $P^{k}$ can be obtained by solving
\begin{equation}\label{inner_PE}
\mathscr{L}_{[{K}_p^{(k,j-1)},{L}_p^{k-1},\mathcal{S}_p]}(P^{(k,j)})+Q_p^{(k,j-1)}=0, 
\end{equation}
where $Q_p^{(k,j)}={Q}_p^{k-1}+|{{K}_p^{(k,j)}}|_{{R}}^{2}$. Subsequently, ${K}_p^{(k,j)}$ is updated by using
\begin{equation}\label{inner_PI}
{K}_p^{(k,j)}=\mathscr{K}_{\mathcal{U}_p}^1(P^{(k,j)}), ~~j=1,2,\cdots.
\end{equation}
\begin{remark}
The inner-loop iteration described above is known as the PI algorithm. This algorithm was originally introduced in \cite{kleinman1968iterative} for deterministic systems and later extended to stochastic systems with multiplicative noise in \cite{li2022stochastic}. Traditional convergence analyses of the PI algorithms rely on the assumptions that the state weighting matrix are at least nonnegative definite. However, in our context, when $k>1$, the updates of $ {L}_p^{k}$ may cause ${Q}_p^{k}$ to be indefinite. Thus, the sufficient conditions for the existing convergence results are no longer applicable. To address this issue, we establish a new convergence result of the PI algorithm that remains valid even when the state weighting matrix is indefinite.
\end{remark}
\begin{theorem}\label{thm:convregence-inner-loop}
 Suppose the conditions of Theorem \ref{thm:convergence-outer-loop} are satisfied and there exists ${K}_p^{(k,0)}\in\mathcal{Z}({L}_p^{k-1})$. {\color{black}Let $P^k\geq0$ and $K_p^k$ be the unique solution of \eqref{outer_eq}.} Let the sequences $\{P^{(k,j)},{K}_p^{(k,j)}\}_{j=1}^{\infty}$ be generated by iteratively solving Eqs. (\ref{inner_PE}) and (\ref{inner_PI}). Then for any $k\in\mathbb{N}_+$, the following properties hold for all $j\in\mathbb{N}_+:$
 \begin{enumerate}
 \item ${K}_p^{(k,j-1)}\in\mathcal{Z}({L}_p^{k-1})$;
 \item $P^{(k,j)}\geq P^{(k,j+1)}\geq P^{k}$;
 \item $\lim_{j\rightarrow\infty}{P}^{(k,j)}={P}^{k}$ and $\lim_{j\rightarrow\infty}{K}_p^{(k,j)}={K}_p^{k}$.
 \end{enumerate}
\end{theorem}
\bproof
For notational simplicity, denote ${R}_p^{(k,j)}\triangleq{R}+D^{\top}P^{(k,j)}{D}$, $\Delta{P}^{(k,j)}\triangleq{P}^{(k,j)}-{P}^k$, $\Delta{K}_p^{(k,j)}\triangleq{K}_p^{(k,j)}-{K}_p^{k}$, $\tilde{P}^{(k,j)}\triangleq{P}^{(k,j)}-{P}^{k(k,j+1)}$, and $\tilde{K}_p^{(k,j)}\triangleq{K}_p^{(k,j)}-{K}_p^{k(k,j-1)}$. 

We first prove that property 1) and the inequality ${P}^{(k,j)} \geq P^{k}$ hold for all $j \geq 1$ by induction on $j$, and subsequently establish the monotonic sequence $\{P^{(k,j)}\}_{j=1}^{\infty}$.

\begin{enumerate}
    \item[\romannumeral1).] For $j=1$, subtracting \eqref{outer_eq:xik+1} from \eqref{inner_PE} and completing the squares yields
    \begin{equation}\label{eq:0814-1}
\mathscr{L}_{[K_{p}^{(k,0)},L_p^{k-1},\mathcal{S}_p]}(\Delta P^{(k,1)})+|\Delta K_p^{(k,0)}|_{R_p^k}^2=0.
    \end{equation}
Since $K_p^{(k,0)}\in\mathcal{Z}(L_p^{k-1})$ holds by assumption, it implies that $\sigma(\mathscr{L}_{[K_p^{(k,0)},L_p^{k-1},\mathcal{S}_p]})\subset\mathbb{C}^-$. By  \cite[Theorem 3.2.3]{sun2020stochastic} and $R_p^k>0$, we conclude $\Delta P^{(k,1)}\geq0$, hence $P^{(k,1)}\geq P^k\geq0$.
\item[\romannumeral2).] Assume that for some $q\geq1$, property 1) holds for $j=q$, we will show that $P^{(k,q)}\geq P^{k}$ and property 1) also holds for $j=q+1$. Eq. (\ref{inner_PE}) at iteration $j=q$ yields
\begin{equation}\label{th2eq2}
\begin{aligned}
\mathscr{L}_{[{K}_p^{(k,q-1)},{L}_p^{k-1},\mathcal{S}_p]}(P^{(k,q)})+ {Q}_p^{k-1}+|{K}_{p}^{(k,q-1)}|_{{R}}^{2}=0.
\end{aligned}
\end{equation}
Subtracting (\ref{outer_eq:xik+1}) from (\ref{th2eq2}) gives
\begin{equation}\label{mathscrL}
\begin{aligned}
\mathscr{L}_{[{K}_p^{(k,q-1)},{L}_p^{k-1},\mathcal{S}_p]}(\Delta{P}^{(k,q)})+|\Delta{K}_p^{(k,q-1)}|_{{R}_p^k}^2\!=\!0.
\end{aligned}
\end{equation}
The same reasoning as in the base case implies $\Delta{P}^{(k,q)}\geq 0$, hence ${P}^{(k,q)}\geq P^k$.

To verify property 1) for $j=q+1$, rewrite the above equation as
\begin{equation}
\begin{aligned}
\mathscr{L}_{[{K}_p^{(k,q)},{L}_p^{k-1},\mathcal{S}_p]}(\Delta{P}^{(k,q)})+{H}_p^{(k,q)}=0,
\end{aligned}
\end{equation}
where ${H}_{(k,q)} =|\Delta{K}_p^{(k,q)}|_{{R}^k_p}^2+|\tilde{{K}}_p^{(k,q)}|_{{R}_p^{(k,q)}}^2\geq0$. Together with $\Delta P^{(k,q)}\geq0$, Proposition \eqref{propo:GLE} ensures that $\sigma(\mathscr{L}_{[K_p^{(k,q)},L_p^{k-1},\mathcal{S}_p]})\subset\mathbb{C}^-$ if $[{A}\!-\!{B}{K}_p^{(k,q)}\!+\!{G}{L}_p^{k-1},{H}_p^{(k,q)}|C-DK_p^{(k,q)}]$ is exactly detectable. Suppose, for contradiction, that this condition does not hold. Then there exists a nonzero $P\in\mathbb{S}^n$ such that 
\begin{equation*}
\mathscr{L}_{[{K}_{p}^{(k,q)},{L}_{p}^{k-1},\mathcal{S}_p]}(P)=\lambda P,~{H}_{p}^{(k,q)}P=0,~\mathrm{Re}(\lambda)\geq0.
\end{equation*} 
Since ${R}_{p}^{k},{R}_{p}^{(k,q)}\!>\!0$, the equality ${H}_p^{(k,q)}P = 0$ leads to
\begin{equation*}
\begin{aligned}
\mathscr{L}^*_{[{K}_{p}^{(k,q-1)},{L}_p^{k-1},\mathcal{S}_{p}]}(P)=\lambda P,~\mathrm{Re}(\lambda)\geq0,
\end{aligned}
\end{equation*}
which contradicts $\sigma(\mathscr{L}_{[{K}_{p}^{(k,q-1)},{L}_p^{k-1},\mathcal{S}_{p}]})\!\subset\!\mathbb{C}^-$, so the exact detectability condition is verified. Hence, $\sigma(\mathscr{L}_{[{K}_{p}^{(k,q)},{L}_{p}^{k-1},\mathcal{S}_p]})\subset\mathbb{C}^-$, thereby ${K}_p^{(k,q)}\in\mathcal{Z}({L}_p^{k-1})$.
\end{enumerate}
Thus, by induction, property 1) and inequality $P^{(k,j)}\geq P^{k}$ hold for all $j\in\mathbb{N}_+$.

From equations (\ref{inner_PE}) and (\ref{inner_PI}), we have
\begin{equation}\label{eq:0924-1}
\begin{aligned}
\mathscr{L}_{[{K}_{p}^{(k,j)},{L}_{p}^{k-1},\mathcal{S}_{p}]}(\tilde{P}^{(k,j)})+|\tilde{{K}}_{p}^{(k,j)}|_{{R}_{p}^{(k,j)}}^{2}=0.
\end{aligned}
\end{equation}
Since $\mathscr{L}_{[{K}_{p}^{(k,j)},{L}_{p}^{k-1},\mathcal{S}_{p}]}\subset \mathbb{C}^-$ has been established for all $j\in\mathbb{N}_+$, Eq. \eqref{eq:0924-1} implies $\tilde{P}^{(k,j)}\geq0$, {\sl i.e.,} $P^{(k,j)}\geq P^{(k,j+1)}$ for all $j$. {\color{black}Therefore, the sequence $\{P^{(k,j)}\}_{j=1}^{\infty}$ is monotonically deceasing and bounded below by $P^{k}$. By the monotone convergence theorem (see \cite[p.189]{kantorovich1964functional}), there exists $P^{(k,\infty)}$ such that $\lim_{j\rightarrow\infty}P^{(k,j)}=P^{(k,\infty)}$. Taking the limit in \eqref{inner_PE} as $j\rightarrow\infty$ gives
\begin{equation*}
\begin{aligned}
&(A+GL_p^{k-1})^{\top}P^{(k,\infty)}\!+\!P^{(k,\infty)}(A+GL_p^{k-1})\!+\!C^{\top}P^{(k,\infty)}C\\
&+Q_p^{k-1}-(P^{(k,\infty)}B+C^{\top}P^{(k,\infty)}D)(R+D^{\top}P^{(k,\infty)}D)^{-1}\\
&\times(B^{\top}P^{(k,\infty)}+D^{\top}P^{(k,\infty)}C)=0.
\end{aligned}
\end{equation*}
Since $P^{k}\geq0$ is the unique stabilizing solution of this equation, we have $\lim_{j\rightarrow\infty}P^{(k,j)}=P^{k}$.} Hence, $\lim_{j\rightarrow\infty}K^{(k,j)}=K^{k}$ by Eq. \eqref{inner_PI}. This completes the proof.
\eproof

To approximate the stabilizing solution $S^k$ of \eqref{outer_eq:ARE}, we replace $P^{(k,j)},K_p^{(k,j-1)},L_p^{k-1},Q,\mathcal{S}_p,\mathcal{U}_p$ in Eqs. \eqref{inner_PE}-\eqref{inner_PI} with $S^{(k,j)},K_s^{(k,j-1)},L_s^{k-1},Q_s,\mathcal{S}_s,\mathcal{U}_s$ to obtain the corresponding inner-loop iteration, where $S^{(k,j)}$ is solved by
\begin{equation}\label{inner_are_PE}
\begin{aligned}
&\mathscr{L}_{[K_s^{(k,j-1)},L_s^{k-1},\mathcal{S}_s]}(S^{(k,j)})+Q_s-\gamma^{2}|L_s^{k-1}|^2\\
&+|K_s^{(k,j-1)}|_{\Upsilon}^2=0,
\end{aligned}
\end{equation}
and $K_s^{(k,j)}$ is updated by
\begin{equation}\label{inner_are_PI}
K_s^{(k,j)}=\mathscr{K}_{\mathcal{U}_s}^1(S^{(k,j)}),~~j=1,2,\cdots.
\end{equation}

{\color{black} An immediate corollary of Theorem \ref{thm:convregence-inner-loop} is stated below.
\begin{corollary}\label{coro:2}
Suppose the conditions of Corollary \ref{coro:1} are satisfied and there exists $K_s^{(k,0)}$ such that $\sigma(\mathscr{L}_{[K_s^{(k,0)},L_s^{k-1},\mathcal{S}_s]})\subset\mathbb{C}^-$. Let $S^k\geq0$ and $K^k$ be the unique solution of \eqref{outer_eq:ARE}. Let the sequences $\{S^{(k,j)},K_s^{(k,j)}\}_{j=1}^{\infty}$ be generated by iteratively solving Eqs. \eqref{inner_are_PE} and \eqref{inner_are_PI}. Then $\lim_{j\rightarrow\infty}S^{(k,j)}=S^k$ and $\lim_{j\rightarrow\infty}K_s^{(k,j)}=K_s^k$.  
\end{corollary}}
{\color{black}\begin{remark}\label{rem:init}
To initialize each inner-loop iteration for both cases, one needs stabilizing gain matrices. Following the approach in \cite[Theorem 1]{rami2000linear}, such gains can be computed via LMIs. Specifically, the initial stabilizers are given by ${K}_p^{(k,0)}=-YX^{-1}$ and ${K}_{s}^{(k,0)}=-VW^{-1}$, which are obtained by solving the LMIs
\begin{equation}\label{eq:LMI1}
\!\!\!\!\!\left[\begin{array}{cc}
\!\!\!{A}_p^{k-1} X\!+\!X ({A}_p^{k-1})^{\!\top}\!+\!{B}Y\!+\!Y^{\top}\! {B}^{\top}&\!\! {C}X\!+\!DY\!\!\\
\!\!\!XC^{\top}\!+\!Y^{\top}\!D&\!\!-X\!\!\!
\end{array}\right] \!<\! -\epsilon I,
\end{equation}
% and
\begin{equation}\label{eq:LMI2}
\left[\begin{array}{cc}
\!\!\!{A}_{s}^{k-1} W\!+\!W ({A}_{s}^{k-1})^{\!\top}\!+\!{B}V\!+\!V^{\top}\!  {B}^{\top}&\!\! 0\!\!\\
\!\!\!0\!&  -W\!\!\!
\end{array}\right] \!<\! -\epsilon I,
\end{equation}
respectively, where ${A}_p^{k-1} = {A}+{G}{L}_p^{k-1}$, ${A}_{s}^{k-1} ={A}_s+G{{L}}_{s}^{k-1}$, and $\epsilon>0$.  
\end{remark}}
 
\section{Robustness Analysis for the Dual-Loop Algorithm}\label{sec:main result2}
In the previous section, we introduced the outer-loop and inner-loop iterations and analyzed their convergence properties under ideal conditions. In this section, we examine the performance of the dual-loop algorithm when only a finite number of iterations is executed and assess their robustness in the presence of external disturbances. 

For the subsequent analysis, we define the sets $\mathcal{D}_{\zeta}\triangleq \{{L}\in\mathcal{W}|\mathrm{Tr}(P^*-P_L)\leq\zeta\}$ and $\mathcal{G}_{\rho}({L})\triangleq \{{K}\in\mathcal{Z}({L})|\mathrm{Tr}(P_K-P_L)\leq\rho\}$,  and conclude some auxiliary results in Appendix \ref{Auxiliary results} based on these formulations. 

The theorems below demonstrate the algorithm’s linear convergence rate. Detailed proofs of these theorems are provided in Appendices \ref{thmpf:rate-outer-loop} and \ref{thmpf:rate-inner-loop}. 
\begin{theorem}\label{thm:rate-outer-loop}
For any $\zeta>0$ and any ${L}_p^{k-1}\in\mathcal{D}_{\zeta}$, there exists  $\alpha \in(0,1)$ such that 
\begin{equation}\label{eq:resul_thm_rate-outer-loop}
\mathrm{Tr}\left(\Delta{P}^{k+1}\right)\leq \alpha \mathrm{Tr}\left(\Delta{P}^k\right).
\end{equation}
\end{theorem}

\begin{theorem}\label{thm:rate-inner-loop}
For any $\rho>0$ and any ${K}_p^{(k,j-1)}\in {\mathcal{G}}_{{\rho}}({L}_p^{k-1})$, there exists $\tilde{\alpha} \in(0,1)$ such that 
\begin{equation}\label{eq:resul_thm_rate-inner-loop}
\mathrm{Tr}\left(\Delta{P}^{(k,j+1)}\right)\leq \tilde{\alpha}\mathrm{Tr}\left(\Delta{P}^{(k,j)}\right).
\end{equation}
\end{theorem}
\begin{remark}
Note that the convergence rate of the inner-loop iteration is independent of ${L}_p^{k-1}$, ensuring that the number of inner-loop iterations does not increase with each outer-loop iterations. Moreover, both theorems provide the bounds $\|\tilde{P}^{k}\|_2\leq(1+\alpha)\alpha^{k-1}\zeta$ and $\|\tilde{P}^{(k,j)}\|_2\leq(1+\tilde{\alpha})\tilde{\alpha}^{j-1}\rho$, with similar bounds for $\tilde{{L}}_p^k$ and $\tilde{{K}}_p^{(k,j)}$. Therefore, a convergence criterion $\xi(k,j)$ can be applied in practice to achieve the desired approximation with a finite number of steps for both loops of the algorithm, as detailed in Algorithm \ref{Alg_1}. {\color{black}Similar results apply to the deterministic iterations \eqref{outer_eq:ARE}-\eqref{outer_eq2:are} and \eqref{inner_are_PE}-\eqref{inner_are_PI}. By substituting them together with the LMI \eqref{eq:LMI2} into Algorithm \ref{Alg_1}, we obtain an approximate solution of Eq. \eqref{are3}.} 
\end{remark}
\begin{algorithm}
  Initialization: $k \leftarrow1$,~$j \leftarrow1$,~${L}_p^0\leftarrow 0$\; 
 \While{$\|\tilde{{L}}_p^{k-1}\|_2 \geq \xi$ or $k=1$}{
 \textcolor{black}{Solve ${K}_p^{(k,0)}$ from Eq. \eqref{eq:LMI1}}\; 
 \While{$\|\tilde{{K}}_p^{(k,j)}\|_2 \geq \xi$ or $j=1$}{
Solve $P^{(k,j)}$ from Eq. \eqref{inner_PE}\;
Update ${K}_p^{(k,j)}$ by Eq. \eqref{inner_PI}\;
$j\leftarrow j+1$ \;
 }
 $P^{k}\leftarrow P^{(k,j)}$\;
 Update ${L}_p^k$ by Eq. \eqref{outer_eq2}\;
  $k\leftarrow k+1$\;
  $j\leftarrow 1$\;
 }
 \caption{Dual-loop Iterative Algorithm}\label{Alg_1}
\end{algorithm}
\begin{remark}
While the convergence criterion in Algorithm \ref{Alg_1} allows to approximate the true solution within a finite number of steps, it also introduces estimation errors. Since each approximation serves as the input for the subsequent iteration, biases may accumulate throughout the iterative process. Additionally, these biases may be further compounded by model mismatches and measurement errors inherent in the data-driven methods. Therefore, it is necessary to analyze the robustness of the iterative processes against these external disturbances.
\end{remark}

Consider the outer-loop iteration at $k\geq1$. A disturbance $\delta{L}_p^k$ may arise, introducing a bias in ${L}_p^{k}$ as follows  
\begin{equation}\label{eq:hat-Lpk}
\hat{{L}}_p^{k} = {L}_p^k+\delta {L}_p^{k}, \quad{L}_p^{k}=\mathcal{K}_{\mathcal{V}}^2(\hat{P}^k).
\end{equation}
This bias leads to an inexact model for the $(k+1)$-th outer-loop iteration, described by
\begin{subequations}\label{hat_outer_eq}
\begin{empheq}[left={\left\{\begin{aligned}},right={\end{aligned}\right.}]{align}
&{\mathscr{L}}_{[{{K}}_p^{k+1},\hat{{L}}_p^{k},\mathcal{S}_p]}(\hat{P}^{k+1})  +  \hat{Q}_p^k + |{{{K}}_p^{k+1}}|_{{R}}^2=0, \\
&{K}_p^{k+1}=\mathcal{K}_{\mathcal{U}_p}^1(\hat{P}^{k+1}),
\end{empheq}
\end{subequations}
where $\hat{{Q}}_p^k={Q}-\gamma^2|\hat{{L}}_p^{k}|^2$. The ``hat'' notation differentiates the exact solution from \eqref{general:oueter-loop} and the inexact one in (\ref{hat_outer_eq}).

Similarly, in the inner-loop iteration at $j\geq1$, a disturbance $\delta{K}_p^{(k,j)}$ is introduced 
\begin{equation}\label{eq:hatkp}
\hat{{K}}_p^{(k,j)}={K}_p^{(k,j)}+\delta{K}_p^{(k,j)}, \quad{K}_p^{(k,j)}=\mathscr{K}_{\mathcal{U}_p}^1(\hat{P}^{(k,j)})
\end{equation} 
This results in inexact iterative equation for the $(j+1)$-th step
\begin{equation}\label{inexact:inner-loop}
\mathcal{L}_{[\hat{{K}}_p^{(k,j)},\hat{{L}}_p^{k-1},\mathcal{S}_p]}(\hat{P}^{(k,j+1)})+\hat{{Q}}_p^{k-1}+|\hat{{K}}_p^{(k,j)}|_{{R}}^2=0.
\end{equation}
\begin{remark}
As illustrated in Fig. \ref{f:diag9}, the inexact iterative equation (\ref{hat_outer_eq}) can be viewed as a discrete-time nonlinear system, where $\hat{P}^k$ is the state and $\delta {L}_p^{k}$ serves as the disturbance input. Similarly, the inexact policy evaluation (\ref{inexact:inner-loop}) can be represented as a discrete-time linear system, where $\hat{P}^{(k,j)}$ is the state and $\delta {K}_p^{(k,j)}$ acts as the disturbance input. In the absence of disturbances, Theorems \ref{thm:rate-outer-loop} and \ref{thm:rate-inner-loop} guarantee that these systems are exponentially stable, and the states $\hat{P}_p^{k}$ and $\hat{P}_p^{(k,j)}$ converges to the equilibria $P^*$ and ${P}^{k}$, respectively. 
\begin{figure}[!htb]
\centering
\includegraphics[scale=0.32]{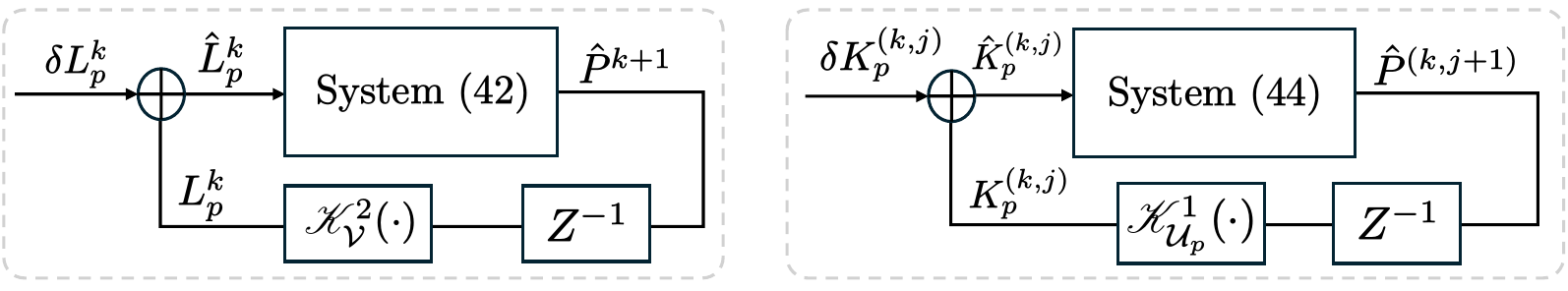}
\centering
\caption{Schematic diagrams of the inexact iterative procedures: (left) inexact outer-loop iteration; (right) inexact inner-loop iteration.}\label{f:diag9}
\end{figure}
\end{remark}

{\color{black}By employing class $\mathcal{K}$ and $\mathcal{KL}$ functions (see \cite[Definition 4.2\&4.3]{khalil2002nonlinear}),} we establish the following robustness theorems to demonstrate the robustness to small disturbances. Their proofs are provided in Appendices \ref{thmpf:robust-outer-loop} and \ref{thmpf:robust-inner-loop}. 

{\color{black}Let  $\zeta^*\!=\!\mathrm{Tr}(P^*)$, $\|\delta {L}_p\|_{\infty}\!=\!\sup_{k\in\mathbb{N}_+}\|\delta {L}_p^k\|_F$, and} {\color{black}$\|\delta{K}_p^k\|_{\infty}\!=\!\sup_{j\in\mathbb{N}_+}\|\delta K_{p}^{(k,j)}\|_F$.}
 
\begin{theorem}\label{thm:robust-outer-loop}
{\color{black}Suppose the conditions of Theorem \ref{thm:convergence-outer-loop} are satisfied. Let $\{\hat{P}^{k},\hat{L}_p^{k}\}_{k=1}^{\infty}$ be generated by iteratively solving \eqref{hat_outer_eq} and \eqref{eq:hat-Lpk} with the initial condition  $\hat{L}_p^{0}=0$. For any $\zeta\geq\zeta^*$, there exists a constant $\varepsilon>0$ such that if the disturbance sequence $\delta {L}_p \triangleq\{\delta {L}_p^{k}\}_{k=1}^{\infty}$ satisfies $\|\delta{L}_p\|_{\infty}<\varepsilon$, then there exist a $\mathcal{K}\mathcal{L}$-function $\ell(\cdot,\cdot)$ and a $\mathcal{K}$-function $\kappa(\cdot)$ such that for all $k\in\mathbb{N}_+$, 
\begin{equation}
\|P^*-\hat{P}^{k+1}\|_F\leq \ell(\|{P}^*-\hat{P}^1\|_F,k)+\kappa(\|\delta{L}_p\|_{\infty}).
\end{equation} }
\end{theorem}  
\begin{theorem}\label{thm:robust-inner-loop}
{\color{black}Suppose the conditions of Theorem \ref{thm:robust-outer-loop} are satisfied. Fix any outer-loop index $k$ and any $\rho>0$. Let $\{\hat{P}_k^{(k,j)},\hat{K}_p^{(k,j)}\}_{j=1}^{\infty}$ be generated by iteratively solving \eqref{inexact:inner-loop} and \eqref{eq:hatkp} with the initial condition $\hat{K}_p^{(k,0)}\in\mathcal{D}_{\rho}(\hat{L}_{p}^{k-1})$. Then there exist  $\tilde{\varepsilon}>0$, a $\mathcal{K}\mathcal{L}$-function $\tilde{\ell}(\cdot,\cdot)$, and a $\mathcal{K}$-function $\tilde{\kappa}(\cdot)$, such that if $\delta{K}_p^k\triangleq\{\delta{K}_p^{(k,j)}\}_{j=1}^{\infty}$ satisfies $\|\delta{K}_p^k\|_{\infty}<\tilde{\varepsilon}$, the following inequality holds for all $j\in\mathbb{N}_+$
\begin{equation}
\!\!\!\!\|\hat{P}^{(k,j)}-{P}^{k}\|_F\leq\tilde{\ell}(\|\hat{P}^{(k,1)}-{P}^{k}\|_F,j)+\tilde{\kappa}(\|\delta {K}_p^k\|_{\infty}). 
\end{equation}}
\end{theorem} 
\begin{remark}
These results ensure that the systems are robust in the sense of small-disturbance ISS (\!\!\cite{pang2021robust}). In practice, when errors occur during the execution of the algorithm, the limit of the approximation error is upper bounded by the magnitude of the disturbances. As a result, the  approximation errors will eventually converge to a small neighborhood around zero, as long as the disturbances remain small.
\end{remark}

\section{Data-driven Implementation of the Iterative Framework for Robust MFSC}\label{sec:main result3}
 {\color{black}
Building on the proposed dual-loop iterative framework, we develop a data-driven method to compute the decentralized control strategy (\ref{uo}) and the decentralized disturbance strategy (\ref{do}) that achieve the asymptotic robust social optimality for the $N$-agent game described by (\ref{sys1}) and (\ref{Jsoc}).

To this end, we first derive data-driven iterative equations for approximating $(P^*,K_p^*,L_p^*)$. Starting from system \eqref{sys1} and applying It\^{o}'s formula to $x_i^{\top}P^{(k,j)}x_i$, we obtain
\begin{equation*}
\begin{aligned}
&\mathrm{d}(x_i^{\top}P^{(k,j)}x_i)=2x_i^{\top}P^{(k,j)}(Ax_i\!+\!Bu_i\!+\!Gv_i)\mathrm{dt}\!+\!(Cx_i\!\\
&+\!Du_i)^{\!\top}P^{(k,j)}(Cx_i+Du_i)\mathrm{d}t+2x_i^{\top} P^{(k,j)}(Cx_i+Du_i)\mathrm{d}w_i.
\end{aligned}
\end{equation*}
Introducing the shorthand variables $M^{(k,j)}\triangleq R_p^{(k,j)}K_p^{(k,j)}$, $L_p^{(k,j)}\triangleq\mathscr{K}_{\mathcal{V}}^2(P^{(k,j)})$, $\Lambda^{(k,j)}\triangleq D^{\top}P^{(k,j)}D$, and $\mu_i^{(k,j)}\triangleq K_p^{(k,j)}x_i$, and substituting the inner-loop Eqs. \eqref{inner_PE}-\eqref{inner_PI}, the expression simplifies to
\begin{equation*}
\begin{aligned}
&\mathrm{d}(x_i^{\top}P^{(k,j)}x_i)=2(K_p^{(k,j-1)}x_i+u_i)^{\top}M^{(k,j)}x_i\mathrm{d}t\\
&-2\gamma^2(L_p^{k-1}x_i-v_i)^{\top}L_p^{(k,j)}x_i\mathrm{d}t-|\mu_i^{(k,j-1)})|_{\Lambda^{(k,j)}}^2\mathrm{d}t\\
&+|u_i|_{\Lambda^{(k,j)}}^2\mathrm{d}t\!-\!|x_i|^2_{Q_p^{(k,j-1)}}\mathrm{d}t\!+\!2x_i^{\top} P^{(k,j)}(Cx_i\!+\!Du_i)\mathrm{d}w_i.
\end{aligned}
\end{equation*}
% \begin{equation*}
% \begin{aligned}
% &\mathrm{d}(x_i^{\top}P^{(k,j)}x_i)=2(K_p^{(k,j-1)}x_i+u_i)^{\top}M^{(k,j)}x_i\mathrm{d}t-2\gamma^2\\
% &\times(L_p^{k-1}x_i-v_i)^{\top}L_p^{(k,j)}x_i\mathrm{d}t-(\mu_i^{(k,j-1)})^{\top}\Lambda^{(k,j)}\mu_i^{(k,j-1)}\mathrm{d}t\\
% &+u_i^{\top}\Lambda^{(k,j)}u_i\mathrm{d}t\!-\!x_i^{\top}Q_p^{(k,j-1)}x_i\mathrm{d}t\!+\!2x_i^{\top} P^{(k,j)}(Cx_i\!+\!Du_i)\mathrm{d}w_i.
% \end{aligned}
% \end{equation*}
Integrating both sides of the above equation along \eqref{sys1} over the interval $[t,t+T]$, taking expectations, and using Kronecker product representation yield the regression form
\begin{equation}\label{eq:irl_sare_vec}
    \psi_{p}^{(k,j-1)}(t)^{ \top} \theta_p^{(k,j)} = -(\bar{I}_{x_i}^{t})^{ \top} \mathrm{vecm}(Q_p^{(k,j-1)}),
\end{equation}
where $\theta_p^{(k,j)}=[\mathrm{vecm}(P^{(k,j)})^{\top}, \mathrm{vec}(M^{(k,j)})^{\top}, \mathrm{vec}(L_p^{(k,j)})^{\top},$ $\mathrm{vecm}(\Lambda^{(k,j)})^{\top}]^{\top}$ and
\[
\psi_p^{(k,j)}(t)=\left[\begin{array}{c}
     \bar{\delta}_{x_i}^t  \\
      - 2\bar{I}_{x_iu_i}^t - 2(I\otimes K_p^{(k,j)})\bar{I}_{x_ix_i}^t\\
      - 2\gamma^2\bar{I}_{x_iv_i}^t + 2\gamma^2(I\otimes L_p^{k-1})\bar{I}_{x_ix_i}^t\\
      - \bar{I}_{u_i}^t + \bar{I}_{\mu_{i}^{(k,j)}}^t
\end{array}\right].
\]
Here, $\bar{\delta}_{{x}_{i}}^t=\mathbb{E}[\delta_{{x}_{{i}}}^t]$ and $\bar{I}_{(\cdot)}^t=\mathbb{E}[I_{(\cdot)}^t]$ with the subscript $(\cdot)$ indicating the corresponding trajectory term.

During the data collection phase $[t_1,t_l]$ with instants $t_k = t_1 + (k-1)T_s$, we construct the following matrices
\begin{equation*}
\begin{aligned}
&\Psi_p^{(k,j)}=\left[\psi_p^{(k,j)}(t_1),\psi_p^{(k,j)}(t_2),\cdots,\psi_p^{(k,j)}(t_l)\right]^{\top},\\
&\bar{\Delta}_{(\cdot)}=\left[\bar{\delta}_{(\cdot)}^{t_1},\bar{\delta}_{(\cdot)}^{t_2},\cdots,\bar{\delta}_{(\cdot)}^{t_l}\right]^{\top},~\bar{\mathcal{I}}_{(\cdot)}=\left[\bar{I}_{(\cdot)}^{t_1},\bar{I}_{(\cdot)}^{t_2},\cdots,\bar{I}_{(\cdot)}^{t_l}\right]^{\top}.
\end{aligned}
\end{equation*}
With these definitions, Eq. \eqref{eq:irl_sare_vec} can be expressed in the  linear matrix form
\begin{equation}
\Psi_p^{(k,j-1)}\theta_p^{(k,j)}=-\bar{\mathcal{I}}_{x_{i}}\mathrm{vecm}(Q_p^{(k,j-1)}).
\end{equation}
Consequently, the $(k,j)$-th approximations of $(P^*,K_p^*,L_p^*)$ can be determined by
\begin{numcases}{}
\theta_p^{(k,j)} =-(\Psi_p^{(k,j-1)} )^{\dag} \bar{\mathcal{I}}_{x_{i}}\mathrm{vecm}(Q_p^{(k,j-1)}),\label{eq:MFK-data-iteration}\\
K_p^{(k,j)}=(R+\Lambda^{(k,j)})^{-1}M^{(k,j)}.\label{eq:MFK-data-iteration2}
\end{numcases}
To guarantee uniqueness of the solution, we impose the following rank condition.
\begin{assume}\label{assume:rank1}
There exists $l_1>0$ such that for all $l\geq l_1$, 
\begin{equation}\label{A:rank1}
\begin{aligned}
&\mathrm{rank}\left(\left[\bar{\mathcal{I}}_{x_{i}},\bar{\mathcal{I}}_{x_{{i}}u_{{i}}},\bar{\mathcal{I}}_{x_{{i}}v_{{i}}},\bar{\mathcal{I}}_{{u}_{{i}}}\right]\right) \\
= &\frac{1}{2}n(n+1)+n(m_1+m_2)+\frac{1}{2}m_1(m_1+1).
\end{aligned}
\end{equation}
\end{assume}
\begin{lemma}
Suppose Assumption \ref{assume:rank1} and the conditions of Theorem \ref{thm:convregence-inner-loop} hold. Let $\{P^{(k,j)},K_p^{(k,j)},L_p^{(k,j)}\}_{j=1}^{\infty}$ be generated by iteratively solving Eqs. \eqref{eq:MFK-data-iteration}-\eqref{eq:MFK-data-iteration2}. Then we have $\lim_{j\rightarrow\infty}P^{(k,j)}=P^k$, $\lim_{j\rightarrow\infty}K_p^{(k,j)}=K_p^k$, and $\lim_{j\rightarrow\infty}L_p^{(k,j)}=L_p^k$.  
\end{lemma}

The proof is similar to that in \cite[Theorem 3.1]{xu2024mean} and is omitted here for brevity. 

Next, we derive a data-driven iterative equation for approximating $(S^*,K_s^*,L_s^*)$. Two main difficulties arise in constructing such equations: 1). the matrix $Q_s$ in \eqref{inner_are_PE} involves system parameters, which prevents complete removal of the dynamical model; 2). the trajectories generated by \eqref{sys1} cannot be directly applied to eliminate system dependence in \eqref{inner_are_PE}-\eqref{inner_are_PI}.

To overcome the first issue, we define the variables 
$$\Pi^{(k,j)} \!\triangleq\! S^{(k,j)}\! -\! P^*,K_{\pi}^{(k,j)}  \!\triangleq \!K_s^{(k,j)}\! -\! \mathscr{K}_{\mathcal{U}_s}^1(P^*),L^{k}_{\pi}\!\triangleq\!L_s^k - L_p^*.$$
Subtracting \eqref{are1} from \eqref{inner_are_PE} yields
\begin{equation}\label{MFPi:iteration-inner-loop-11} 
\mathscr{L}_{[K_{\pi}^{(k,j-1)},L^{k-1}_{\pi},\mathcal{S}_{\pi}]}(\Pi^{(k,j)})\!+\!Q_{\Gamma}^{k-1}\!+\!|K_{\pi}^{(k,j-1)}|_{\Upsilon}^2\!=\!0,
\end{equation}
where $$\mathcal{S}_{\pi}=[A-BK_p^*+GL_p^*,B,G|0,0], ~Q_{\Gamma}^k=-Q_{\Gamma}-\gamma^2|{L}_{\pi}^{k}|^2.$$
Moreover, the updates of $K_{\pi}^{(k,j)}$ and $L_{\pi}^{(k,j)}$ are given by
\begin{align}
&{K}_{\pi}^{(k,j)} = \mathscr{K}^1_{\mathcal{U}_s}(\Pi^{(k,j)}),\label{MFPi:iteration-inner-loop-12}\\
&{L}_{\pi}^{(k,j)}=\mathscr{K}^2_{\mathcal{V}}(\Pi^{(k,j)}). \label{MFPi:iteration-outer-loop-22}
\end{align}
Compared with \eqref{inner_are_PE}, the matrix $-Q_{\Gamma}$ of Eq. \eqref{MFPi:iteration-inner-loop-11} eliminates the dependence on the system parameters. Moreover, by Corollary 2 and the existence of the unique stabilizing solution $P^*\geq0$ to \eqref{are1}, the convergence of the iteration between \eqref{MFPi:iteration-inner-loop-11} and \eqref{MFPi:iteration-inner-loop-12} is guaranteed.

Furthermore, to address the second issue, we introduce the expected state, control, and disturbance variables
\begin{equation}\label{Exuv:eq}
\bar{x}_{{i}}(t) = \mathbb{E}[x_{{i}}(t)],~\bar{u}_{{i}}(t) = \mathbb{E}[u_{{i}}(t)],~\bar{v}_{{i}}(t) = \mathbb{E}[v_{{i}}(t)],
\end{equation}
which satisfy $\mathrm{d}\bar{x}_{{i}}=(A\bar{x}_{{i}}+B\bar{u}_{{i}}+G\bar{v}_{{i}})\mathrm{d}t.$

From \eqref{MFPi:iteration-inner-loop-11}-\eqref{MFPi:iteration-outer-loop-22}, the time derivative of $\bar{x}_i^{\top}\Pi^{(k,j)}\bar{x}_i$ is obtained as
\begin{equation*}
\begin{aligned}
&\frac{\mathrm{d}}{\mathrm{d}t}(\bar{x}_i^{\top}\Pi^{(k,j)}\bar{x}_i)= 2(\bar{u}_i+K_{\pi}^{(k,j-1)}\bar{x}_i)^{\top}\Upsilon K_{\pi}^{(k,j)}\bar{x}_i+2\gamma^2(\bar{v}_i\\
&-L_{\pi}^{k-1}\bar{x}_i)^{\top}L_{\pi}^{(k,j)}\bar{x}_i-\bar{x}_i^{\top}(Q_{\Gamma}^{k-1}+|K_{\pi}^{(k,j-1)}|_{\Upsilon}^2)\bar{x}_i.
\end{aligned}
\end{equation*}
Integrating it over $[t,t+T]$ and using Kronecker product representation yields
\begin{equation}\label{eq:irl2_vec}
\psi_{\pi}^{(k,j-1)}(t)^{\top}\theta_{\pi}^{(k,j)}=-(I_{\bar{x}_i}^t)^{\top}\mathrm{vecm}(Q_{\Gamma}^{k-1}+|K_{\pi}^{(k,j-1)}|_{\Upsilon}^2),
\end{equation}
where $\theta_{\pi}^{(k,j)}=[\mathrm{vecm}(\Pi^{(k,j)})^{\top},\mathrm{vec}(K_{\pi}^{(k,j)})^{\top},\mathrm{vec}(L_{\pi}^{(k,j)})^{\top} ]^{\top}$,
\[\psi_{\pi}^{(k,j)}\! =\! \left[\begin{array}{c}
    \!\!\!\delta_{\bar{x}_i}^t\!\!\!\\
    \!\!\!-2(I\!\otimes\! \Upsilon)I_{\bar{x}_i\bar{u}_i}^t\!-\!2(I\!\otimes\! \Upsilon(K_p^*\!+\!K_{\pi}^{(k,j-1)}))I_{\bar{x}_i\bar{x}_i}^t\!\!\!\\
\!\!\!-2\gamma^2I_{\bar{x}_i\bar{v}_i}^t+2\gamma^2(I\otimes(L_p^*+L_{\pi}^{k-1}))I_{\bar{x}_i\bar{x}_i}^t\!\!\!
    \end{array}\right].
\]
Define the matrices over $[t_1,t_l]$ as
\begin{equation*}
\begin{aligned}
&\Psi_{\pi}^{(k,j)}=\left[\psi_{\pi}^{(k,j)}(t_1),\psi_{\pi}^{(k,j)}(t_2),\cdots,\psi_{\pi}^{(k,j)}(t_l)\right]^{\top},\\
&{\Delta}_{(\cdot)}=\left[{\delta}_{(\cdot)}^{t_1},{\delta}_{(\cdot)}^{t_2},\cdots,{\delta}_{(\cdot)}^{t_l}\right]^{\top},~{\mathcal{I}}_{(\cdot)}=\left[{I}_{(\cdot)}^{t_1},{I}_{(\cdot)}^{t_2},\cdots,{I}_{(\cdot)}^{t_l}\right]^{\top}.
\end{aligned}
\end{equation*}
 Eq. \eqref{eq:irl2_vec} can be written in matrix form as
 \begin{equation}
\Psi_{\pi}^{(k,j-1)}\theta_{\pi}^{(k,j)}=-{\mathcal{I}}_{\bar{x}_{i}}\mathrm{vecm}(Q_{\Gamma}^{k-1}+|K_{\pi}^{(k,j-1)}|_{\Upsilon}^2).
 \end{equation}
Hence, the $(k,j)$-th approximation of $(\Pi^*,K_{\pi}^*,L_{\pi}^*)$ is
\begin{equation}\label{eq:MFK-data-iteration_pi}
\begin{aligned}
\theta_{\pi}^{(k,j)} =-(\Psi_{\pi}^{(k,j-1)} )^{\dag} {\mathcal{I}}_{\bar{x}_{i}}\mathrm{vecm}(Q_{\Gamma}^{k-1}\!+\!|K_{\pi}^{(k,j-1)}|_{\Upsilon}^2). 
\end{aligned}
\end{equation} 
\begin{assume}\label{assume:rank2}
There exists $l_2>0$ such that for all $l\geq l_2$,
\begin{equation}\label{A:rank2}
\begin{aligned}
\mathrm{rank}\left(\left[\mathcal{I}_{\bar{x}_{{i}}},\mathcal{I}_{\bar{x}_{{i}}\bar{u}_{{i}}},\mathcal{I}_{\bar{x}_{{i}}\bar{v}_{{i}}}\right]\right) \!=\! \frac{1}{2}n(n\!+\!1)\!+\!n(m_1\!+\!m_2).
\end{aligned}
\end{equation}
\end{assume}

\begin{lemma}
Suppose Assumption \ref{assume:rank2} and the conditions of Corollary \ref{coro:2} hold. Let $\{P^{(k,j)},K_p^{(k,j)},L_p^{(k,j)}\}_{j=1}^{\infty}$ be generated by iteratively solving Eqs. \eqref{eq:MFK-data-iteration}-\eqref{eq:MFK-data-iteration2}. Then we have $\lim_{j\rightarrow\infty}\Pi^{(k,j)}=S^k-P^*$, $\lim_{j\rightarrow\infty}K_{\pi}^{(k,j)}=K_s^k-\mathscr{K}_{\mathcal{U}_s}^1(P^*)$, and $\lim_{j\rightarrow\infty}L_{\pi}^{(k,j)}=L_s^k-L_p^*$.  
\end{lemma}

The proof is similar to that in \cite[Theorem 3.2]{xu2024mean}} {\color{black} and is therefore omitted.

To fulfill the rank conditions 
\eqref{A:rank1} and \eqref{A:rank2}, exploratory control and disturbance strategies are designed as
\begin{align*}
&u_{{i}}(t) \!=\! -K_{exp}x_{{i}}(t)+\xi^1(t),~ \xi^1(t)\!=\!\sigma_1\sum\nolimits_{j=1}^{n_1}\sin(\omega_jt),\\
&v_{{i}}(t) \!=\! L_{exp}x_{{i}}(t)+\xi^2(t),~ \xi^2(t)\!=\!\sigma_2\sum\nolimits_{j=1}^{n_2}\sin(\upsilon_jt),
\end{align*}
where $K_{exp},L_{exp}$ are exploratory gains and $\xi^1,\xi^2$ are typically used as the probing noise in IRL algorithms \cite{jiang2012computational,xu2024mean}. 

In practice, we estimate the expected quantities using a finite number of trajectory samples as $\hat{\bar{(\cdot)}}=\frac{1}{N_s}\sum_{j}^{N_s}(\cdot)^j$, where $N_s$ is the number of samples and the superscript $j$ represents the $j$-th sample. By the law of large numbers, as $N_s\rightarrow\infty$, these estimates converge to their true values. Substituting these approximations into (\ref{eq:MFK-data-iteration})-\eqref{eq:MFK-data-iteration2} and (\ref{eq:MFK-data-iteration_pi}), the undetermined parameters are solved by
\begin{numcases}{}
\hat{\theta}_{p}^{(k,j)}=-(\hat{\Psi}_{p}^{(k,j-1)})^{\dag}\hat{\bar{\mathcal{I}}}_{x_{i}}\mathrm{vecm}(\hat{Q}_p^{(k,j-1)}),\label{eq:MFK-data-iteration-hat}\\
K_p^{(k,j)}=(R+\Lambda^{(k,j)})^{-1}M^{(k,j)},\label{eq:MFK-data-iteration2-hat}
\end{numcases}
\begin{equation}\label{eq:MFK-data-iteration_pi-hat}
\begin{aligned}
\!\hat{\theta}_{\pi}^{(k,j)}=-(\hat{\Psi}_{\pi}^{(k,j-1)})^{\dag}{\mathcal{I}}_{\hat{\bar{x}}_{i}}\mathrm{vecm}(\hat{Q}_{\Gamma}^{k-1}+|\hat{K}_{\pi}^{(k,j-1)}|_{\Upsilon}^2).
\end{aligned}
\end{equation}
Here, the hats denote data-based approximations and the corresponding estimated parameters. 
\begin{remark}
As stated in Remark \ref{rem:init}, computing the initial stabilizing gain matrices via the LMI approach requires knowledge of the system parameters. One option is to identify these parameters. In particular, by introducing the quadratic term  $\bar{x}_{{i}}^{\top}E_j\bar{x}_i$ and following the same derivation steps as in the data-driven iterations, the $j$-th rows of matrices $A,B,G$, denoted by $A_j,B_j,G_j$, can be reconstructed as $[A_j,B_j,G_j]^{\top}=\Phi_j^{\dag}\Delta_{\bar{x}_i}\mathrm{vecm}(E_j)$. Here, $\Phi_j  =2\left[\mathcal{I}_{\bar{x}_{{i}}\bar{x}_{{i}}}(e_j\!\otimes\! I),\mathcal{I}_{\bar{x}_{{i}}\bar{u}_{{i}}}(e_j\!\otimes\! I),\mathcal{I}_{\bar{x}_{{i}}\bar{v}_{{i}}}(e_j\!\otimes\! I)\right]$,  $E_j\in\mathbb{R}^{n\times n}$ and $e_j\in\mathbb{R}^n$ denote the matrix and vector whose $j$-th diagonal element and $j$-th entry are equal to one, respectively, while all the other entries are zero. With $(A,B,G)$ identified and $(C,D)$ available, the LMIs in Remark \ref{rem:init} an then be solved to yield admissible stabilizers. Alternatively, a bootstrap PI-based IRL method \cite{Chen2025boot} can be employed to directly find stabilizing gains from input–state data, without requiring any prior knowledge of the system dynamics.
\end{remark}}

Finally, to compute the mean field state $\bar{x}(t)$, we set  the control input and disturbance input to 
\begin{equation}
u_{i}=-(\hat{K}_p^*+\hat{K}_{\pi}^*)x_{{i}},~~v_i=(\hat{L}_p^{*}+\hat{L}_{\pi}^*)x_i
\end{equation}
with $\hat{K}_p^*,\hat{K}_{\pi}^*,\hat{L}_p^*,\hat{L}_{\pi}^*$ are obtained from the iterative procedure. Here, ${K}_{\pi}^*=K_s^*-\mathcal{K}_{\mathcal{U}_s}^1(P^*)$, $L_p^*=L_s^*-L_p^*$, and the hats denote the corresponding estimates.

By \eqref{xbar}, the mean field approximation can be computed as
\begin{equation}
\hat{\bar{x}}(t)\!=\!\frac{1}{N_s}\sum\nolimits_{j=1}^{N_s}\!{x}_{{i}}^j(t),~~t\geq0,
\end{equation}
where $x_{{i}}^j$ represents the $j$-th state sample of $\mathcal{A}_{{i}}$.

\section{Numerical example}\label{sec:simulation}
In this section, we present a numerical simulation to validate the effectiveness of the proposed algorithm. The large-scale population consists of $500$ agents, with the system parameters
\begin{equation} \label{sys:simulation}
\begin{aligned}
&A=\left[\begin{array}{cc}
0.3&0.7\\
-0.9&0.5
\end{array}\right],~B= \left[\begin{array}{c}
0.2\\
0
\end{array}\right],~G=\left[\begin{array}{c}
0.1\\
0
\end{array}\right],\\
&C=\left[\begin{array}{cc}
0.01&0.03\\
0.05&0.02
\end{array}\right],~D=\left[\begin{array}{cc}
0.05\\
0.05
\end{array}\right].
\end{aligned}
\end{equation} 
The initial state $x_{i0}$ is uniformly distributed on $[-4,0]\times[0,4]\subset\mathbb{R}^2$ with $\mathbb{E}[x_{i0}]=[-2,2]^{\top}$. 

The coefficients of the cost function (\ref{Jsoc}) are given by 
\begin{equation*}
\begin{aligned}
\!\!\!\!\!\!Q=\left[\begin{array}{cc}
10&0\\
0&10
\end{array}\right], ~\Gamma=\left[\begin{array}{cc}
0.9&0\\
0&0.9
\end{array}\right], ~R=1.25, ~\gamma = 2.
\end{aligned}
\end{equation*}
 
The data-driven method for approximating decentralized strategies begins by collecting measurements from $\mathcal{A}_i$. Its control and disturbance inputs are designed as 
\begin{equation}\label{u12}
\begin{aligned}
&K_{exp}=[6,-3],~~\xi^1(t)=5\sum\nolimits_{j=1}^{100}\sin(w_jt),\\ 
&L_{exp}=[0,0],~~\xi^2(t)=10\sum\nolimits_{j=1}^{100}\!\sin(\upsilon_jt),
\end{aligned}
\end{equation}
where $w_j\in[-100,100]$ and $\upsilon_j\in[-300,300]$ are randomly selected. The other parameters are set as follows: $T_s=0.001$ [sec], $T=0.1$ [sec], $[t_1,t_l]=[0,14]\text{~[sec]}$, and the convergence criterion $\xi=10^{-5}$. The scalar on the right-hand side of the LMIs (\ref{eq:LMI1}) and (\ref{eq:LMI2}) are set to $\epsilon=5$.

\begin{figure*}[!htb]
\centering
\includegraphics[scale=0.36]{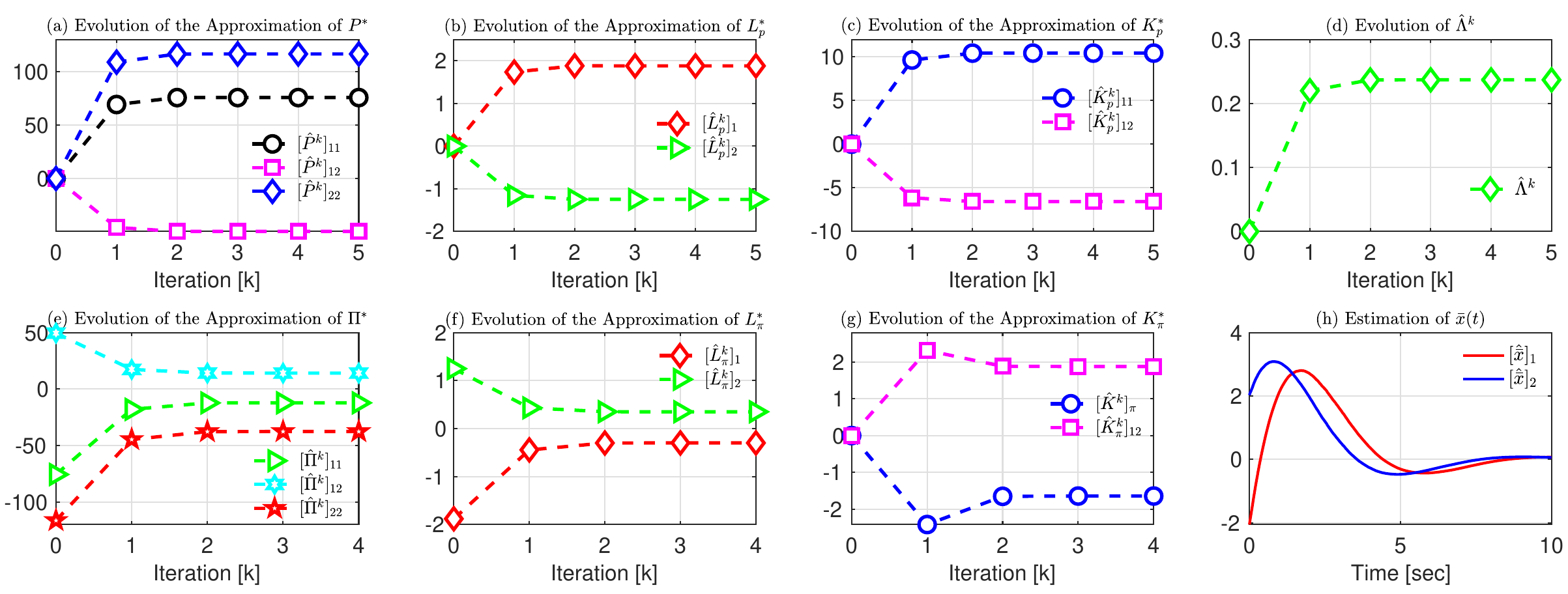}
\centering
\caption{Determined parameters of the data-driven method.}\label{f:rf1}
\end{figure*}

The rank conditions (\ref{A:rank1}) and (\ref{A:rank2}) are verified using sample paths from the system (\ref{sys:simulation}) with (\ref{u12}) at the end of the data collection phase. The first iterative process then starts with the estimation of system parameters and stabilizing control inputs. Under the convergence criterion $\varepsilon$, the inner-loop iterations converge, and the final values are used to update $\{\hat{P}^k,\hat{L}_p^k,\hat{K}_p^k,\hat{\Lambda}^k\}$. The results of the outer-loop iterations are shown in Fig. \ref{f:rf1} (a)-(d), demonstrating convergence at the $4$-th iteration based on $\xi=10^{-5}$. The second iterative process is then applied to obtain the sequences $\{\hat{\Pi}^k,\hat{L}_{\pi}^k,\hat{K}^k_{\pi}\}$, which also converge at the $4$-th iteration, as illustrated in Fig. \ref{f:rf1} (e)-(g). 

\begin{figure*}[!htb]
\centering
\includegraphics[scale=0.4]{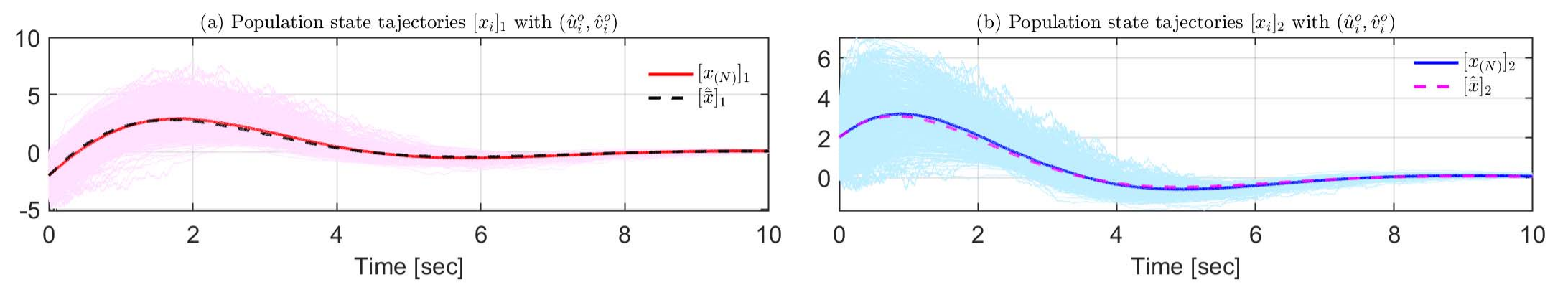}
\centering
\caption{State trajectories of the population with $(\hat{u}_p^o,\hat{v}_p^o)$.}\label{f:rf2}
\end{figure*}

\begin{table*}[!htb]
\begin{center}
  \begin{tabular}{|@{\  \hspace*{3mm}}l||*{7}{c|}}\hline
\multicolumn{1}{|@{}l||}{\backslashbox[0pt][l]{Iteration}{Errors}}
&\makebox[4em]{$\frac{\|\hat{P}^k-P^k\|_2}{\|P^k\|_2}$}&\makebox[4em]{$\frac{\|\hat{L}_p^k-L_p^k\|_2}{\|L_p^k\|_2}$}
&\makebox[4em]{$\frac{\|\hat{K}^k_p-K^k_p\|_2}{\|K^k_p\|_2}$}&\makebox[4em]{$\frac{\|\hat{\Lambda}^k-\Lambda^k\|_2}{\|\Lambda^k\|_2}$}&\makebox[4em]{$\frac{\|\hat{\Pi}^k-\Pi^k\|_2}{\|\Pi^k\|_2}$}&\makebox[4em]{$\frac{\|\hat{L}_{\pi}^k-{L}_{\pi}^k\|_2}{\|{L}_{\pi}^k\|_2}$}&\makebox[4em]{$\frac{\|\hat{K}_{\pi}^k-{K}_{\pi}^k\|_2}{\|{K}_{\pi}^k\|_2}$}\\\hline\hline
$k=1$ &$0.0023$& $0.0085$ & $0.0088$& $0.0107$& $0.0091$&$0.0129$&$0.0099$\\\hline
$k=2$ &$0.0047$& $0.0096$ & $0.0073$&$0.0044$&$0.0168$&$0.0172$&$0.0175$\\\hline
$k=3$ &$0.0050$& $0.0104$ &$0.0069$&$0.0040$&$0.0168$&$0.0173$&$0.0175$\\\hline
$k=4$ &$0.0050$& $0.0104$ &$0.0069$&$0.0040$&$0.0168$&$0.0173$&$0.0175$\\\hline
\end{tabular}
\end{center}
\caption{ Estimation errors of the data-driven method.}\label{e1tab1}
\end{table*}

To verify the robustness of the data-driven method, we compare the results with one from Algorithm \ref{Alg_1}, denoted by $\{P^k,L_p^k,K_p^k,\Lambda^k,\Pi^k,L_{\pi}^k,K_{\pi}^k\}$ and summarize the estimation errors in Table. \ref{e1tab1}. Note that both model-based dual-loop iterations are convergent at the $4$-th iteration under the convergence criterion $\xi=10^{-5}$. The small error values indicate that the approximations from the data-driven method closely match the model-based results. It demonstrates that despite minor disturbances introduced by the data-driven estimations, the inner loop still converges to values within a small range of the true values. Furthermore, the convergence of outer-loop iterations under a given convergence criterion highlights the uniform convergence of the overall iterative processes.
 
Then, the control strategy $\hat{u}_{{i}}=-(\hat{K}_p^*+\hat{K}_{\pi}^*)x_{{i}}$ and the disturbance strategy $\hat{v}_{{i}}=(\hat{L}^*_p+\hat{L}^*_{\pi})x_{{i}}$ are applied to drive $\mathcal{A}_{{i}}$. The approximated mean field state trajectories are obtained by averaging the state sample paths, as shown in Fig. \ref{f:rf1} (h).  

Finally, the mean field social optimum is approximated as 
\begin{numcases}{}
\hat{u}_i^o = -\hat{K}^*_px_i(t)-\hat{K}^*_{\pi}\hat{\bar{x}}(t),\\
\hat{v}_i^o = \hat{L}^*_px_i(t)+\hat{L}^*_{\pi}\hat{\bar{x}}(t),~1\leq i\leq500,
\end{numcases}
and are used across the population. \textcolor{black}{The resulting  state trajectories} are shown in Fig. \ref{f:rf2}. The average state trajectories of the population closely match the approximated mean field state trajectories, thereby verifying the consistency condition. 

\section{Conclusion} \label{sec:conclusion}
In the LQG setting, the solution of robust MFSC problem is reduced to solving indefinite Riccati-type equations that govern the gain matrices and an ordinal differential equation that describes the evolution of the mean field state. This reduction enables us to compute the robust social optimum strategies by developing a data-driven method. Notably, the proposed dual-loop framework is robust against disturbances encountered during the iterative process, and the uniform convergence of the overall algorithm is rigorously proven. Moreover, this framework is not only applicable for achieving the asymptotic robust social optimality, but also for solving other robust control or $H_{\infty}$ control problems that involve these types of Riccati equations as part of the optimal solution. {\color{black}As a direction for future work, we plan to extend this study to the more general case where coupling appears not only in the} {\color{black}cost functional but also explicitly in the agents’ dynamics. This introduces additional analytical challenges for developing data-driven methods.}
 
\appendices  % for no appendix heading
% do not use \section anymore after \appendix, only \section*
% is possibly needed
% use appendices with more than one appendix
% then use \section to start each appendix
% you must declare a \section before using any
% \subsection or using \label (\appendices by itself
% starts a section numbered zero.)
\section{Proof of Lemma \ref{lem:used in Theorem1}}\label{pflem:used in Theorem1}
For notational simplicity, let $R_L = R+D^{\top}P_LD$, $H_1 =\gamma^2|L_p^*-L'|^2$, $H_2 = \gamma^2|L-L'|^2$, and $H_3 = |{K}_p^* - {K}_{L}|_{R_L}^2$.

By using the operator $\mathscr{L}_{[K_p^*,L',\mathcal{S}_p]}$ and completing the squares, we can rewrite equations (\ref{are_operator1}) and \eqref{general:oueter-loop1} as  
\begin{align}
&\mathscr{L}_{[K_p^*,L',\mathcal{S}_p]}(P^*)+H_1+Q_{L}=0,\label{eq:lem1_xi}\\
&\mathscr{L}_{[K_{p}^*,L',\mathcal{S}_p]}(P_L)\!-\!H_2\!-H_3+\!{Q}_L\!=\!0,\label{eq:lem1_xik}
\end{align}
respectively, where ${Q}_L={Q}-\gamma^2|L'|^2+|{K}_p^*|_{R}^2$.
Subtracting Eq. \eqref{eq:lem1_xik} from Eq. \eqref{eq:lem1_xi} yields
\begin{equation}\label{eq:lem1_Deltaxik}
\begin{aligned}
\mathscr{L}_{[{K}_p^*,L',\mathcal{S}_p]}(P^*-{P}_{L})+H=0,
\end{aligned}
\end{equation}
where $H=H_1+H_2+H_3$.
Since $L\in\mathcal{W}$ implies that $P^*-P_L\geq0$ and $H\geq0$, to apply Proposition \ref{propo:GLE} to conclude $\sigma(\mathscr{L}_{[K_p^*,L',\mathcal{S}_p]})\subset\mathbb{C}^-$, the rest is to prove that $[A-BK_p^*+GL',H|C-DK_p^*]$ is exactly detectable. 

We prove by contradiction. Assume the system is not exactly detectable, then there exists a nonzero $P\in\mathbb{S}^n$ such that 
\begin{equation}\label{eq1:pflemm1}
\mathscr{L}_{[K_p^*,L',\mathcal{S}_p]}(P)=\lambda P,~~HP=0,~~\mathrm{Re}(\lambda)\geq0.
\end{equation}
Since $R_L>0$ and $\gamma^2>0$, the equation $HP=0$ implies that $L_p^*P=L'P$. Substituting this equation into \eqref{eq1:pflemm1} yields
\begin{equation*}
\mathscr{L}_{[K_p^*,L_p^*,\mathcal{S}_p]}(P)=\lambda P,~~\mathrm{Re}(\lambda)\geq0,
\end{equation*}
which contradicts the fact $\sigma(\mathscr{L}_{[K_p^*,L_p^*\mathcal{S}_p]})\subset\mathbb{C}^-$. Therefore, the exact detectability condition is satisfied. As a result, we immediately conclude that $\sigma(\mathscr{L}_{[K_p^*,L',\mathcal{S}_p]})\subset\mathbb{C}^-$.\hfill $\square$

\section{Auxiliary results}\label{Auxiliary results}
\begin{lemma}\label{thm3:lem2}
$P_{\!L}(L)$ is continuous for any $L\in\!\mathcal{W}$.
\end{lemma}
\bproof
Define a function $\mathcal{F}:{R}^{m_2\times n}\times \mathbb{S}^n\mapsto\mathbb{S}^n$ by the left-hand side of Eq. \eqref{general:oueter-loop}, which can be written as
\begin{equation*}
\mathcal{F} =\mathcal{F}_1(L,P_L)+\mathcal{F}_2(P_L)+{Q}-\gamma^2|{L}|^2,
\end{equation*}
where $\mathcal{F}_1 =  ({A} + {G}{L})^{ \top} P_L + P_L({A} + {G}{L}) + {C}^{\top} {P}_L{C}$, $\mathcal{F}_2  =  -(P_L {B} + {C}^{ \top} P_L{D})(R + D^{ \top} P_L D)^{-1}  ({B}^{ \top} P_L + {D}^{ \top} P_L{C})$.
By using the Kronecker product, the derivative of $\mathrm{vec}(\mathcal{F}_1)$ with respect to $\mathrm{vec}(P_L)$ can be expressed as
\begin{equation}\label{fi1}
\frac{\partial \mathrm{vec}(\mathcal{F}_1)}{\partial \mathrm{vec}(P_L)} \!=\! \left( ({A}\!+\!{G}{L}) \!\otimes \!I\!+\!I\!\otimes\!({A}\!+\!{G}{L})\!+\!{C}\!\otimes\!{C}\right)^{\!\top}\!.
\end{equation}
Based on the matrix inverse derivative identity and  Eq. \eqref{general:oueter-loop2}, the differential of $\mathcal{F}_2$ can be computed by
\begin{equation*}
\begin{aligned}
\mathrm{d}\mathcal{F}_2  =  &-(\mathrm{d} P_L{B} + {C}^{\top} \mathrm{d}P_L{D}){K}_L + {K}_L^{\top} {D}^{\top}\mathrm{d}P_L{D}{K}_{{L}}\\
&-{K}_{L}^{\top}( {B}^{\top}\mathrm{d}P_L + {D}^{\top} \mathrm{d}P_LC).
\end{aligned}
\end{equation*}
Vectorizing $\mathrm{d}\mathcal{F}_2$ to express it in terms of $\mathrm{d}\mathrm{vec}(P_L)$ yields
\begin{equation}\label{fi2}
\begin{aligned}
\frac{\mathrm{d}\mathrm{vec}(\mathcal{F}_2)}{\mathrm{d}\mathrm{vec}(P_L)} &\! = \!-\!\Big(({B}{K}_{{L}})\!\otimes\!I \!+\! I\!\otimes\!({B}{K}_{{L}})\!+\!({D}{K}_{{L}})\!\otimes\!{C}\\
&+ {C}\!\otimes\!({D} {K}_{{L}})\!-\!({D} {K}_{{L}})\!\otimes\!({D} {K}_{{L}}) \Big)^{\top}.
\end{aligned}
\end{equation}
Eqs. \eqref{fi1} and (\ref{fi2}) together imply that the Jacobian matrix satisfies $\frac{\partial \mathrm{vec}(\mathcal{F})}{\partial \mathrm{vec}(P_L)}= {M}$, where ${M}= (({A}+{G}{L}_p-{B}{K}_{{L}})\otimes I+I\otimes ({A}+{G}{L}-{B}{K}_{{L}})+({C}-{D}{K}_{{L}})\otimes ({C}-{D}{K}_{{L}}))^{\top}$.
 
Since ${L}\in\mathcal{W}$, we have $\sigma(\mathscr{L}_{[{K}_{{L}},{L},\mathcal{S}_p]})\subset \mathbb{C}^-$, which implies that the matrix ${M}$ is Hurwitz. Consequently, the Jacobian matrix $\frac{\partial \mathrm{vec}(\mathcal{F})}{\partial \mathrm{vec}(P_L)}$ is invertible. By the implicit function theorem \cite{krantz2002implicit}, there exists a continuously differentiable function $P_L({L})$ such that $\mathcal{F}({L},P_L({L}))=0$. Hence, $P_L$ is continuous with respect to ${L}$ for any ${L}\in\mathcal{W}$.
\eproof
\begin{lemma}\label{thm3:lem4}
Let $\zeta>0$ and $\rho>0$. For all matrices ${L}\in\mathcal{D}_{\zeta}$ and ${K}\in{\mathcal{G}}_{\rho}({L})$, consider the solution $\{\Phi(t),t\geq0\}$ to the matrix SDE
\begin{equation*} 
\mathrm{d}\Phi   =  ({A} - {B}{K} + {G}{L})\Phi\mathrm{d}t + ({C} - {D}{K})\Phi\mathrm{d}{{w}},~\Phi(0)=I.
\end{equation*}
Define ${P}=\mathbb{E}\int_{0}^{\infty}\Phi(t)\Phi(t)^{\top}\mathrm{d}t$. Then there exists $0<\underline{b}\leq\overline{b}<\infty$ such that 
\begin{equation}
\underline{b}\leq \lambda_{\min}({P}) \leq \mathrm{Tr}({P}) \leq \overline{b}.
\end{equation}  
\end{lemma}

\bproof 
Let $S(t)\! =\! \mathbb{E}\!\left[\Phi(t)\Phi(t)^{\!\top}\right]$. By It\^{o}'s formula, it gives
\begin{equation*} 
\frac{{\mathrm{d}}S(t)}{{\mathrm{d}t}}=\mathscr{L}_{[{K},{L},\mathcal{S}_p]}^*(S(t)),~~S(0)=I.
\end{equation*}
Since $\sigma(\mathscr{L}_{[K,L,\mathcal{S}_p]})\subset\mathbb{C}^-$, we have $S(\infty)=0$. Hence, ${P}=\int_0^{\infty}S(\tau)\mathrm{d}\tau$ satisfies the generalized Lyapunov equation
\begin{equation} \label{sys:P}
\mathscr{L}_{[{K},{L},\mathcal{S}_p]}({P}) +I=0.
\end{equation}
By \cite[Theorem 3.2.3]{sun2020stochastic}, this equation admits a unique positive definite solution $P$.

Furthermore, both sets $\mathcal{D}_{\zeta}$ and ${\mathcal{G}}_{{\rho}}({L})$ can be shown to be closed and bounded, and thus compact by the Heine–Borel theorem (\!\!\cite[Theorem 2.41]{rudin1964principles}). The joint admissible set $\{(L,K)|L\in\mathcal{D}_{\zeta},K\in\mathcal{Z}(L)\}$ is also compact. Since the mapping $(L,{K})\mapsto {P}$ defined in \eqref{sys:P} is continuous, it follows from the extreme value theorem \cite{rudin1964principles} that there exist uniform constants $0<\underline{b}\leq\overline{b}<\infty$, independent of ${L}$ and ${K}$, such that $\underline{b}\leq \lambda_{\min}({P}) \leq \mathrm{Tr}({P}) \leq \overline{b}$. 
\eproof

\section{Proof of Theorem \ref{thm:rate-outer-loop}}\label{thmpf:rate-outer-loop}
We begin by showing that for any $\zeta>0$ and any ${L}_p^{k-1}\in\mathcal{D}_{\zeta}$, there exists $a >0$ such that 
\begin{equation}\label{a(delta)}
\|\Delta P^k\|_F\leq a \|\Xi_p^k\|_F,
\end{equation}
where $\Xi_p^k = \gamma^2|\tilde{{L}}_p^{k-1}|^2$.
 
To establish this, subtract  (\ref{outer_eq:xik+1}) from (\ref{are_operator1}) to obtain
\begin{equation*}
\begin{aligned}
&\mathscr{L}_{[{K}_p^*,{L}_p^*,\mathcal{S}_p]}(\Delta{P}^k) - \gamma^2|\Delta{L}_p^k|^2+ \Xi_p^k + |\Delta{K}_p^{k}|^2_{{R}_p^{k}}=0.
\end{aligned}
\end{equation*}
Since $\sigma(\mathscr{L}_{[{K}_p^*,{L}_p^*,\mathcal{S}_p]})\subset\mathbb{C}^-$, we have
\begin{equation*}
\Delta P^k\!=\!\mathbb{E}\!\!\int_0^{\infty}\!\!\Phi^*(t)^{\!\top}\!\! \left(\!-\gamma^2|\Delta{L}_p^k|^2\!+\!\Xi_p^k\!+\!|\Delta{K}_p^{k}|^2_{{R}_p^{k}}\right) \!\Phi^*(t)\mathrm{d}t,
\end{equation*}
where $\Phi^*(t)$ satisfies the matrix SDE
\begin{equation*} 
\mathrm{d}\Phi^* = ({A} - {B}{K}_p^* + {G}{L}_p^*)\Phi^*\mathrm{d}t + ({C} - {D}{K}_p^*)\Phi\mathrm{d}{{w}},~\Phi(0)=I.
\end{equation*}
Discard the term $-\gamma^2|\Delta{L}_p^k|^2\leq0$ and take the trace. Due to $\mathrm{Tr}(XY)\leq \mathrm{Tr}(X)\mathrm{Tr}(Y)$ for $X,Y\geq0$, we know
\begin{equation*}
\begin{aligned}
\mathrm{Tr}(\Delta P^k)\leq \mathrm{Tr}(P_{\Phi^*})\mathrm{Tr}(\Xi_p^k)+\mathrm{Tr}(P_{\Phi^*})\mathrm{Tr}(|\Delta{K}_p^k|^2_{{R}_p^k}),
\end{aligned}
\end{equation*}
where $P_{\Phi^*}=\mathbb{E}\int_{t}^{\infty}\Phi^*(t)\Phi^*(t)^{\top}\mathrm{d}t$. By $L^*\in\mathcal{D}_{\zeta}$, $K_p^*\in\mathcal{G}_{\rho}(L_p^*)$, and Lemma \ref{thm3:lem4}, we have $\mathrm{Tr}(P_{\Phi^*})\leq\overline{b}$, and thus
\begin{equation}\label{0928-2}
\mathrm{Tr}(\Delta P^k)\leq \overline{b}\mathrm{Tr}(\Xi_p^k)+\overline{b}\mathrm{Tr}\left(|\Delta{K}_p^k|^2_{{R}_p^k}\right).
\end{equation}
Proceeding similarly to the derivation of \eqref{ktilde}, we can get
\begin{equation*} 
\begin{aligned}
\Delta{K}_p^k=\Upsilon^{-1}({B}^{\top}\Delta{P}^k+{D}^{\top}\Delta{P}^k{C}_p^k),
\end{aligned}
\end{equation*}
which combined with \cite[Lemma 1]{wang1986trace} gives $$\mathrm{Tr}\left(|\Delta{K}_p^k|_{{R}_p^k}^2\right)\leq \beta_1^k \|\Delta {P}^k\|_2\mathrm{Tr}(\Delta {P}^k) $$ with $\beta_1^k =\|{\Upsilon}^{-1}\|_2^2\|{R}_p^{k}\|_2\left(\|{B}\|_2+\|{C}_p^k\|_2\|{D}\|_2\right)^2$. 
Since ${R}_p^{k}$ and ${C}_p^k$ are continuous with respect to ${P}^{k}$, and Lemma \ref{thm3:lem2} ensures that $L_p^{k-1}\mapsto {P}^{k}$ is continuous on the compact set $\mathcal{D}_{\zeta}$, there exist a constant $b_1>0$ such that $\beta_1^k \leq b_1$ for all ${L}_p^{k-1}\in\mathcal{D}_{\zeta}$. Plugging it into (\ref{0928-2}), it gives 
\begin{equation*} 
\mathrm{Tr}(\Delta P^k)\leq\overline{b}\mathrm{Tr}(\Xi_p^k)+\overline{b} b_1\|\Delta{P}^k\|_2\mathrm{Tr}(\Delta{P}^k).
\end{equation*}
Let $\eta = \frac{1}{2\overline{b}b_1}$. If $\|\Delta {P}^k\|_2\leq \eta $, one has
\begin{equation}\label{ineq1:Xi_p^k}
\|\Delta {P}^k\|_F\leq \mathrm{Tr}(\Delta P^k)\leq 2\overline{b} \mathrm{Tr}(\Xi_p^k)\leq 2\overline{b}\sqrt{n}\|\Xi_p^k\|_F.
\end{equation}
If $\|\Delta {P}^k\|_2\geq\eta$, since $\mathcal{D}_{\zeta}$ is compact, the subset $\mathcal{D}_{\zeta}\cap\{{L}_p^{k-1}\in\mathcal{W}|\|\Delta{P}^k\|_2\geq \eta \}$ is also compact. 
Next we prove $\Xi_p^k\neq0$ by contradiction. Suppose $\Xi_p^k=0$. Since $\gamma^2>0$, this would imply ${L}_p^k={L}_p^{k-1}$. From Eqs. \eqref{outer_eq} and \eqref{outer_eq2}, it then follows that the solution $P^{k}\geq0$ also satisfies Eq. \eqref{are_operator}. By the uniqueness of the solution to \eqref{are_operator}, we obtain ${P}^k=P^*$, which yields $\Delta {P}^k=0$. This contradicts the condition $\|\Delta {P}^k\|_2\geq\eta$. Hence, $\Xi_p^k\neq0$, and therefore 
$\|\Xi_p^k\|_F\neq0$. Finally, by the extreme value theorem \cite{rudin1964principles}, there exists $\epsilon>0$ such that $\|\Xi_p^k\|_F\geq \epsilon$. Thus,
 \begin{equation}\label{ineq2:Xi_p^k}
 \|\Delta {P}^k\|_F\leq \mathrm{Tr}(\Delta {P}^k)\leq \frac{\zeta}{\epsilon}\|\Xi_p^k\|_F.
 \end{equation}
Combining the two cases and defining $a = \max(2\overline{b}\sqrt{n},\frac{\zeta}{\epsilon})$, Eqs. (\ref{ineq1:Xi_p^k}) and (\ref{ineq2:Xi_p^k}) together imply $$\|\Delta {P}^k\|_F \leq a\|\Xi_p^k\|_F,~\forall {L}_p^{k-1}\in\mathcal{D}_{\zeta}.$$

We now show that \eqref{eq:resul_thm_rate-outer-loop} holds for all $L_p^{k-1}\in\mathcal{D}_{\zeta}$.  From (\ref{th1eq7}), we have
\begin{equation}\label{0915-1}
\mathscr{L}_{[{{K}}_p^{k+1},{L}_p^{k},\mathcal{S}_p]}(\tilde{P}^{k})+\Xi_p^{k}+|\tilde{{K}}_p^k|_{{R}_p^k}^2=0.
\end{equation}
Since ${L}_p^{k-1}\in\mathcal{D}_{\zeta}\subseteq\mathcal{W}$, Lemma \ref{lemma:W} implies that $\sigma(\mathscr{L}_{[{K}_p^{k+1},{L}_p^{k},\mathcal{S}_p]})\subset\mathbb{C}^{-}$. Moreover, discard the term $|\tilde{{K}}_p^k|_{{R}_p^k}^2\geq0$, apply the trace operator, and use \cite[Lemma 1]{wang1986trace} together with Lemma \ref{thm3:lem4}. Eq. \eqref{0915-1} then yields
\begin{equation}\label{ineq:Xi_k-1}
\mathrm{Tr}(\tilde{P}^{k})\geq {\underline{b}}\|\Xi_p^{k}\|_F.
\end{equation}
Combining \eqref{a(delta)} and \eqref{ineq:Xi_k-1} gives
\begin{equation}\label{92}
\begin{aligned}
\mathrm{Tr}(\Delta{P}^{k+1})\leq \left(1-\frac{\underline{b}}{\sqrt{n}a}\right)\mathrm{Tr}(\Delta{P}^{k}).
\end{aligned}
\end{equation}
Set $\alpha = 1- \frac{\underline{b}}{\sqrt{n}a}$. Then Eq. (\ref{eq:resul_thm_rate-outer-loop}) is satisfied. Moreover, since $a\geq 2\overline{b}\sqrt{n}$ and $0<\underline{b}/\overline{b}<1$, we have $\frac{\underline{b}}{\sqrt{n}a}\in(0,1)$, which implies $\alpha\in(0,1)$. \hfill $\square$

\section{Proof of Theorem \ref{thm:rate-inner-loop}}\label{thmpf:rate-inner-loop}
Let $\Xi_p^{(k,j)} = |\tilde{{K}}_p^{(k,j)}|_{{R}_p^{(k,j)}}^2$. Using $\mathscr{L}_{[{K}_p^k,{L}_p^{k-1},\mathcal{S}_p]}$, and completing the squares, Eq. \eqref{inner_PE} becomes
\begin{equation*} 
\begin{aligned}
&\mathscr{L}_{[{K}_p^k,{L}_p^{k-1},\mathcal{S}_p]} (P^{(k,j)}) + \Xi_p^{(k,j)} - | \Delta {K}_p^{(k,j)} |_{{R}_p^{(k,j)}} \\
&~~~~+ {Q}_p^{k-1} + | {K}_p^k |_{{R}}^2 = 0.
\end{aligned}
\end{equation*}
Subtracting Eq. \eqref{are_operator1} from the above equation yields
\begin{equation}\label{eq1} 
\mathscr{L}_{[{K}_p^k,{L}_p^{k-1},\mathcal{S}_p]}(\Delta{P}^{(k,j)}) + \Xi_p^{(k,j)}-|\Delta{K}_p^{(k,j)}|_{{R}_p^{(k,j)}}^2 = 0.
\end{equation}
By Lemma \ref{thm3:lem4} and ${R}_p^{(k,j)}>0$, it follows from  (\ref{eq1}) that
\begin{equation*}
\mathrm{Tr}\left(\Delta{P}^{(k,j)}\right) \leq \overline{b}\|\Xi_p^{(k,j)}\|_F.
\end{equation*}
Since ${K}_p^{(k,j-1)}\in\mathcal{G}_{\rho}({L}_p^{k-1})\subseteq\mathcal{Z}({L}_p^{k-1})$ ensures ${K}_p^{(k,j)}\in\mathcal{Z}({L}_p^{k-1})$, we can derive similarly from (\ref{eq:0924-1}) that
\begin{equation*}
\mathrm{Tr}\left(\tilde{P}^{(k,j)}\right)\geq \underline{b}\|\Xi_p^{(k,j)}\|_F.
\end{equation*}
Hence, $
\mathrm{Tr}(\Delta{P}^{(k,j+1)})\leq(1-\underline{b}/\overline{b})\mathrm{Tr}(\Delta{P}^{(k,j)})$.
By setting $\tilde{\alpha}=1-\underline{b}/\overline{b}\in(0,1)$, the proof is complete.\hfill $\square$

\section{Proof of Theorem \ref{thm:robust-outer-loop}}\label{thmpf:robust-outer-loop}
Before proving Theorem \ref{thm:robust-outer-loop}, we establish an important intermediate result.
\begin{lemma}\label{lem:invariant set}
For any $\zeta>0$, let $\hat{{L}}_p^{k-1}\in\mathcal{D}_{\zeta}$ and $\hat{{L}}_p^{k} ={L}_p^{k}+\delta {L}_p^{k} $. Then there exists $\varepsilon>0$ such that $\hat{{L}}_p^{k}\in\mathcal{D}_{\zeta}$ for all $\|\delta {L}_p^{k}\|_F\leq\varepsilon$.
\end{lemma}
\bproof
Since $\hat{{L}}_p^{k-1}\in\mathcal{D}_{\zeta}\subseteq\mathcal{W}$, it follows from Lemma \ref{lemma:W} that ${L}_p^{k}\in\mathcal{W}$. We will show that $\hat{{L}}^k_p\in\mathcal{D}_{\zeta}$. By (\ref{hat_outer_eq}), we have
\begin{equation*} 
\begin{aligned}
\mathscr{L}_{[{K}_{i}^{k+1},\hat{{L}}_p^k,\mathcal{S}_p]}(\hat{P}^{k+1} - \hat{P}^k) + |\tilde{{K}}_p^{k}|_{{R}}^2 + \hat{\Xi}_p^{k} -\gamma^2 |\delta{L}_p^{k}|^2 = 0,
\end{aligned}
\end{equation*}
where $\hat{\Xi}_p^k=\gamma^2|{L}_p^{k}-\hat{{L}}_p^{k-1}|^2$.

Assume $\hat{{L}}_{i}^{k}\in\mathcal{W}$. By applying the same reasoning as in the derivation from \eqref{0915-1} to \eqref{ineq:Xi_k-1}, the above equation yields
\begin{equation}
\begin{aligned}
\mathrm{Tr}(\hat{P}^{k+1}-\hat{P}^{k})\geq \underline{b}\|\hat{\Xi}_p^k\|_F-\overline{b}\gamma^2\|\delta {L}_p^k\|_F^2,
\end{aligned}
\end{equation}
Since $\hat{{L}}_p^{k-1}\in\mathcal{D}_{\zeta}$, the proof of Theorem \ref{thm:rate-outer-loop} gives
\begin{equation*}
\|P^*-\hat{P}^k\|_F\leq a \|\hat{\Xi}_p^k\|_F,
\end{equation*}
where $a=\max(2\bar{b}\sqrt{n},\frac{\zeta}{\epsilon})$. Hence,
\begin{subequations} 
 \begin{empheq}[left={\begin{aligned}},right={\end{aligned}}]{align}
\mathrm{Tr}(P^* - \hat{P}^{k+1})& \leq  \alpha \mathrm{Tr}(P^*-\hat{P}^{k})+\overline{b} \gamma^2\|\delta {L}_p^k\|_F^2\label{eq:0916-1}\\
& \leq  (1 - {\underline{b}}/(\sqrt{n}a))\zeta+\overline{b}\gamma^2\|\delta {L}_p^k\|_F^2.
\end{empheq}
\end{subequations}

Set $\varepsilon= (\frac{\underline{b}\zeta}{\sqrt{n}\overline{b}\gamma^2a})^{\frac{1}{2}}$. This ensures that $\mathrm{Tr}(P^*-\hat{P}^{k+1})\leq \zeta$ whenever $\|\delta {L}_p^k\|_F\leq \varepsilon$. Hence, $\hat{{L}}_p^{k}\in\mathcal{D}_{\zeta}$.
 
Next, we verify that $\hat{{L}}_p^k\in\mathcal{W}$. Suppose, for contradiction, that $\hat{{L}}_p^k\notin\mathcal{W}$. From the preceding argument, the set $\mathcal{W}$ contains every ${L}\in\mathbb{R}^{m_2\times n}$ satisfying $\|{L}-{L}_p^{k}\|_F\leq\varepsilon$. This would imply $\|\delta {L}_p^{k}\|_F\geq\varepsilon$, which contradicts the assumption $\|\delta{L}_p^{k}\|_F\leq\varepsilon$. Therefore, $\hat{{L}}_p^k\in\mathcal{W}$. 
\eproof

\textbf{Proof of Theorem \ref{thm:robust-outer-loop}}  
{\color{black}We first prove, by mathematical induction, that  $\hat{L}_p^{k-1}\in\mathcal{D}_{\zeta}$ holds for all $k\in\mathbb{N}_+$ whenever $\|\delta L_p\|_{\infty}\triangleq \sup_{k\in\mathbb{N}_+}\|\delta L_p^k\|_{F}\leq\varepsilon$, where $\varepsilon= (\frac{\underline{b}\zeta}{\sqrt{n}\overline{b}\gamma^2a})^{\frac{1}{2}}$.

1) For $k=1$, with $\hat{{L}}_p^0=0$, Eq. \eqref{outer_eq} at $k=0$ becomes
\begin{equation}\label{eq:0901-1}
\mathscr{L}_{[K_p^1,0,\mathcal{S}_p]}(\hat{P}^{1})+Q+|K_p^{1}|_R^2=0,~K_p^1=\mathscr{K}_{\mathcal{U}_p}^1(\hat{P}^{k+1}).
\end{equation}
Under Assumption \ref{assume:3-1}, this equation admits a unique solution $\hat{P}^1\geq0$ such that $\sigma(\mathscr{L}_{[K_p^1,0,\mathcal{S}_p]})\subset\mathbb{C}^-$. Subtract \eqref{eq:0901-1} from \eqref{are1}, use $\mathscr{L}_{[K_p^*,0,\mathcal{S}_p]}$, and complete the squares. We have
\begin{equation*}
\mathscr{L}_{[K_p^*,0,\mathcal{S}_p]}(P^*- \hat{P}^1)+\gamma^2|L_p^*|^2+|\Delta K_p^1|_{R_p^1}^2=0.
\end{equation*}
Since $\sigma(\mathscr{L}_{[K_p^*,0,\mathcal{S}_p]}) \subset \mathbb{C}^-$ can be easily verified and $R>0$, it follows that $P^*\geq \hat{P}^1$. Thus, $\mathrm{Tr}(P^* - \hat{P}^1) \leq \zeta^*$. Because $\zeta \geq \zeta^*$, we conclude $\hat{L}_1^{0}\in\mathcal{D}_{\zeta}$. Applying Lemma \ref{lem:invariant set} with $\|\delta L_p\|_{\infty}\leq\varepsilon$ then gives $\hat{L}_p^1\in\mathcal{D}_{\zeta}$.

2) Assume for some $q>1$ that $\hat{L}_{p}^{q-1}\in\mathcal{D}_{\zeta}$. If $\|\delta L_p\|_{\infty}\leq\varepsilon$,  Lemma \ref{lem:invariant set} ensures that $\hat{L}_p^{q}\in\mathcal{D}_{\zeta}$.

By induction, $\hat{L}_p^{k-1}\in\mathcal{D}_{\zeta}$ holds for all $k\in\mathbb{N}_+$ whenever $\|\delta L_p^k\|_{\infty} \triangleq \sup_{k\rightarrow\infty}\|\delta L_p^k\|_{F}\leq\epsilon$. Iterating (\ref{eq:0916-1}) from $k=1$ yields $\mathrm{Tr}(P^* - \hat{P}^{k+1})\leq\alpha^{k}\mathrm{Tr}({P}^* - \hat{{P}}^1) +  \frac{\overline{b}\gamma^2}{\underline{b}}\sqrt{n}a \|\delta {L}_p\|_{\infty}^2$,
which implies $$\|{P}^* - \hat{{P}}^{k+1}\|_F \leq \alpha^{k}\sqrt{n}\|{P}^*-\hat{{P}}^1\|_F  +  \frac{\overline{b}\gamma^2}{\underline{b}}\sqrt{n}a\|\delta {L}_p\|_{\infty}^2.$$ 
By Theorem \ref{thm:rate-outer-loop}, $\alpha\in(0,1)$. Hence, $ \ell(\|{P}^*-\hat{{P}}^1\|_F,k) \triangleq \alpha^{k}\sqrt{n}\|{P}^*-\hat{{P}}^1\|_F$ is a $\mathcal{K}\mathcal{L}$-function, and $\kappa(\|\delta {L}_p\|_{\infty})\triangleq \frac{\overline{b}\gamma^2}{\underline{b}}\sqrt{n}a\|\delta {L}_p\|_{\infty}^2$ is a $\mathcal{K}$-function.  \hfill $\square$}
% \eproof

\section{Proof of Theorem \ref{thm:robust-inner-loop}}\label{thmpf:robust-inner-loop}

\begin{lemma}\label{lem:set-inner-loop}
Given $\hat{{L}}_p^{k-1}\in\mathcal{D}_{\zeta}$. For any $\rho>0$, let $\hat{{K}}_p^{(k,j-1)}\in{\mathcal{G}}_{\rho}(\hat{{L}}_p^{k-1})$ and $\hat{{K}}_p^{(k,j)} = {K}_p^{(k,j)}+\delta {K}_p^{(k,j)}$. There exists  $\tilde{\varepsilon} >0$ such that $\hat{{K}}_p^{(k,j)}\in\mathcal{G}_{\rho}(\hat{{L}}_p^{k-1})$ for all $\|\delta{K}_p^{(k,j)}\|_F\leq\tilde{\varepsilon}$.
\end{lemma}

\bproof
Since $\hat{{K}}_p^{(k,j-1)}\in{\mathcal{G}}_{\rho}(\hat{{L}}_p^{k-1})\subseteq{\mathcal{Z}}(\hat{{L}}_p^{k-1})$, Theorem \ref{thm:convregence-inner-loop} ensures that ${{K}}_p^{(k,j)}\in{\mathcal{Z}}(\hat{{L}}_p^{k-1})$. Assume $\hat{{K}}_{i}^{(k,j)}\in\mathcal{Z}(\hat{{L}}_p^{k-1})$. Then, from Eq. \eqref{inexact:inner-loop}, we have
\begin{equation*}
\begin{aligned}
\!\!\!\mathscr{L}_{[\hat{{K}}_p^{(k,j)},\hat{{L}}_p^{k-1},\mathcal{S}_p]}(\hat{{P}}^{(k,j)}\!\!-\!\!\hat{{P}}^{(k,j+1)})\!+\!\hat{\Xi}_p^{(k,j)}\!\!-\!\!|\delta {K}_p^{(k,j)}|^2_{\hat{{R}}_p^{(k,j)}}\!=\!0,
\end{aligned}
\end{equation*}
where $\hat{\Xi}_p^{(k,j)}=|{K}_p^{(k,j)}-\hat{{K}}_p^{(k,j-1)}|^2_{\hat{{R}}_p^{(k,j)}}$ and $\hat{{R}}_p^{(k,j)} = {R}+{D}^{\top}\hat{{P}}^{(k,j)}{D}$. This implies
\begin{equation*}
\mathrm{Tr}\left(\hat{{P}}^{(k,j)}-\hat{{P}}^{(k,j+1)}\right)\geq \underline{b}\|\hat{\Xi}_p^{(k,j)}\|_F-\overline{b}b_2\|\delta {K}_p^{(k,j)}\|_F^2,
\end{equation*}
where $b_2=\sup_{\hat{K}_p^{(k,j-1)}\in\mathcal{D}_{\rho}(\hat{L}_p^{k-1})}\|\hat{{R}}_p^{(k,j)}\|_2$. As a result, 
\begin{equation}\label{itereq11}
\begin{aligned}
\mathrm{Tr}(\hat{{P}}^{(k,j+1)} - \hat{{P}}^{k}) \leq  \tilde{\alpha}\mathrm{Tr}(\hat{{P}}^{(k,j)} - \hat{{P}}^{k}) + \overline{b}b_2\|\delta {K}_p^{(k,j)}\|_F^2.
\end{aligned}
\end{equation}
Since $\hat{{K}}_p^{(k,j-1)}\in {\mathcal{G}}_{\rho}(\hat{{L}}_p^{k-1})$ and $\tilde{\alpha}=1- \underline{b}/\overline{b}$, \eqref{itereq11} yields
\begin{equation*}
\begin{aligned}
\|\hat{{P}}^{(k,j+1)}-\hat{{P}}_p^{k}\|_F\leq(1- \underline{b}/\overline{b})\rho+\overline{b}b_2\|\delta {K}_p^{(k,j)}\|_F^2.
\end{aligned}
\end{equation*}
Let $\tilde{\varepsilon}(\rho)=(\frac{\rho\underline{b}}{\overline{b}^2b_2})^{\frac{1}{2}}$.
Then $\mathrm{Tr}(\hat{{P}}^{(k,j+1)}-\hat{{P}}^k)\leq \rho$ holds whenever $\|\delta {K}_p^{(k,j)}\|_F\leq\tilde{\varepsilon}(\rho)$, which implies $\hat{{K}}_p^{(k,j)}\in{\mathcal{G}}_{\rho}(\hat{{L}}_p^{k-1})$.

Next, we confirm that $\hat{{K}}_p^{(k,j)}\in{\mathcal{Z}}(\hat{{L}}_p^{k-1})$. Suppose, for contradiction, that $\hat{{K}}_p^{(k,j)}\notin {\mathcal{Z}}(\hat{{L}}_p^{k-1})$. From the above argument, $ {\mathcal{Z}}(\hat{{L}}_p^{k-1})$ contains every ${K}\in\mathbb{R}^{m_1\times n}$  such that $\|{K}-{K}_p^{(k,j)}\|_F\leq \tilde{\varepsilon}$. This would imply $\|\delta {K}_p^{(k,j)}\|_F\geq\tilde{\varepsilon}$, which contradicts the assumption that $\|\delta {K}_p^{(k,j)}\|_F\leq \tilde{\varepsilon}$.
\eproof

{\bf Proof of Theorem \ref{thm:robust-inner-loop}}. 
{\color{black}Given $\hat{{L}}_p^{k-1}\in\mathcal{D}_{\zeta}$, We first show by mathematical induction that $\hat{K}_p^{(k,j-1)}\in\mathcal{G}_{\rho}(\hat{L}_p^{k-1})$ holds for all $j\in\mathbb{N}_+$ whenever $\|\delta K_{p}^k\|_{\infty}\triangleq \sup_{j\in\mathbb{N}_+}\|\delta K_p^{(k,j)}\|_F\leq \tilde{\varepsilon}$, where $\tilde{\varepsilon}(\rho)=(\frac{\rho\underline{b}}{\overline{b}^2b_2})^{\frac{1}{2}}$.

1) For $j=1$, since $\hat{{K}}_p^{(k,0)}\in\mathcal{G}_{\rho}(\hat{{L}}_p^{k-1})$ and $\|\delta K_{p}^{k}\|_{\infty}\leq \tilde{\varepsilon}$, Lemma \ref{lem:set-inner-loop} ensures that $\hat{{K}}_p^{(k,1)}\in\mathcal{G}_{\rho}(\hat{{L}}_p^{k-1})$.

2) Assume for some $q>1$, $\hat{K}_p^{(k,q-1)}\in\mathcal{G}_{\rho}(\hat{{L}}_p^{k-1})$. Since $\|\delta K_{p}^{k}\|_F\leq \tilde{\varepsilon}$, Lemma \ref{lem:set-inner-loop} implies $\hat{K}_p^{(k,q)}\in\mathcal{G}_{\rho}(\hat{{L}}_p^{k-1})$.

By induction, $\hat{K}_p^{(k,j-1)}\in\mathcal{G}_{\rho}(\hat{{L}}_p^{k-1})$ whenever $\|\delta K_{p}^{k}\|_{\infty}\leq \tilde{\varepsilon}$. Repeated application of \eqref{itereq11} from $j=1$ yields
\begin{equation*}
\mathrm{Tr}(\hat{P}^{(k,j+1)} - \hat{{P}}^{k})\leq \tilde{\alpha}^{j}\mathrm{Tr}(\hat{{P}}^{(k,1)} - \hat{{P}}^{k}) + \frac{\overline{b}^2}{\underline{b}}b_2\|\delta{K}_p^k\|_{\infty}^2,
\end{equation*}
which further gives
\begin{equation*}
\|\hat{{P}}^{(k,j+1)} - \hat{{P}}^{k}\|_F\leq \sqrt{n}\tilde{\alpha}^{j}\|\hat{{P}}^{(k,1)} - \hat{{P}}^{k}\|_F  +  \frac{\overline{b}^2}{\underline{b}}b_2\|\delta{K}_p^k\|_{\infty}^2.
\end{equation*}
By Theorem \ref{thm:rate-inner-loop}, we have $\tilde{\alpha}\in(0,1)$. Hence, $\tilde{\ell}(\|\hat{{P}}^{(k,j)} - \hat{{P}}^{k}\|_F,j) \triangleq \sqrt{n}\tilde{\alpha}^{j}\|\hat{{P}}^{(k,1)} - \hat{{P}}^{k}\|_F $ is a $\mathcal{K}\mathcal{L}$-function and $\tilde{\kappa}(\|\delta{K}_p^k\|_{\infty}) \triangleq \frac{\overline{b}^2}{\underline{b}}b_2\|\delta{K}_p^k\|_{\infty}^2$ is a $\mathcal{K}$-function.\hfill $\square$}
%

%% use section* for acknowledgment
%\section*{Acknowledgment}
%
%
%The authors would like to thank...

% Can use something like this to put references on a page
% by themselves when using endfloat and the captionsoff option.
% \ifCLASSOPTIONcaptionsoff
%   \newpage
% \fi

% trigger a \newpage just before the given reference
% number - used to balance the columns on the last page
% adjust value as needed - may need to be readjusted if
% the document is modified later
%\IEEEtriggeratref{8}
% The "triggered" command can be changed if desired:
%\IEEEtriggercmd{\enlargethispage{-5in}}

% references section

% can use a bibliography generated by BibTeX as a .bbl file
% BibTeX documentation can be easily obtained at:
% http://mirror.ctan.org/biblio/bibtex/contrib/doc/
% The IEEEtran BibTeX style support page is at:
% http://www.michaelshell.org/tex/ieeetran/bibtex/
 \bibliographystyle{IEEEtran}
% % argument is your BibTeX string definitions and bibliography database(s)
 \bibliography{autosam_TAC.bib}
\vspace{-1cm}

\end{document}